\documentclass[a4paper,12pt]{iopart}
\usepackage[utf8]{inputenc}
\usepackage{hyperref}
\usepackage{color}
\usepackage{tensor}
\usepackage{ulem}
\eqnobysec 
\newcommand{\eqend}[1]{\,\mathrm{#1}}

\newcommand{\eqref}[1]{(\ref{#1})}
\newcommand{\bra}[1]{{\left\langle{#1}\right\vert}}
\newcommand{\ket}[1]{{\left\vert{#1}\right\rangle}}

\newcommand{\poisson}[2]{\left\lbrace #1,  #2 \right\rbrace _{P}}
\newcommand{\dirac}[2]{\left\lbrace #1, #2 \right\rbrace _{D}}
\newcommand{\fpartial}[2]{\frac{ \partial #1}{\partial #2}}

\newcommand{\anticom}[2]{\left[ #1,  #2 \right]_{+}}
\newcommand{\com}[2]{\left[ #1,  #2 \right]_{-}}
\newcommand{\braket}[2]{\langle #1\! \mid \!#2 \rangle}

\makeatletter

\newcommand\fulllabel[1]{
\def\@currentlabel{\ifnumbysec\arabic{section}.\arabic{eqnval}\else\arabic{eqnval}\fi}
\label{#1}%
\let\@currentlabel\theequation
}

\let\@@ppendixstar\@appendixstar
\def\@appendixstar{\@@ppendixstar
\def\numparts{\addtocounter{equation}{1}%
     \setcounter{eqnval}{\value{equation}}%
     \setcounter{equation}{0}%
     \def\theequation{\ifnumbysec
     \Alph{section}.\arabic{eqnval}{\it\alph{equation}}%
     \else\Alph{section}\arabic{eqnval}{\it\alph{equation}}\fi}}
\def\endnumparts{\def\theequation{\ifnumbysec
     \Alph{section}.\arabic{equation}\else
     \Alph{equation}\fi}%
     \setcounter{equation}{\value{eqnval}}}
}

\makeatother

\usepackage{iopams}

\begin{document}

\title[Supergravity on a 3-torus]{Supergravity on a 3-torus: quantum linearization instabilities with a supergroup}

\author{Atsushi Higuchi\(^1\) and Lasse Schmieding\(^1\)}
\address{\(^1\) Department of Mathematics, University of York, Heslington, York, YO10 5DD, United Kingdom}

\eads{\mailto{atsushi.higuchi@york.ac.uk}, \mailto{lcs549@york.ac.uk} }

\begin{abstract}
It is well known that linearized gravity in spacetimes with compact Cauchy surfaces and continuous symmetries
suffers from linearization instabilities: solutions to classical linearized gravity in such
a spacetime must satisfy so-called linearization stability conditions (or constraints) for them to extend to solutions in the
full non-linear theory.
Moncrief investigated implications of these conditions in linearized quantum gravity in such background spacetimes and
found that
the quantum linearization stability constraints lead to the requirement that all physical states must be invariant under the
symmetries generated by these constraints.
He studied these constraints for linearized quantum gravity in flat spacetime with the spatial sections of
toroidal topology in detail.  Subsequently,
his result was reproduced by the method of group-averaging.
In this paper the quantum linearization stability conditions are studied for $\mathcal{N}=1$ simple supergravity in this spacetime.  In addition to the
linearization stability conditions
corresponding to the spacetime symmetries, i.e.\ spacetime translations, there are also fermionic linearization
stability conditions corresponding to the
background supersymmetry.  We construct all states satisfying these quantum linearization stability conditions, 
including the fermionic ones, and show
that they are obtained by group-averaging over the supergroup of the global supersymmetry of this theory.
\end{abstract}

\maketitle

\section{Introduction}\label{intro}

\noindent In physics, the equations of interest are frequently non-linear and difficult to solve. Typically only a small number of
solutions are known and these exploit special
 symmetries. To extract further physics from the known solutions, a common strategy is to perturbatively expand small
deviations around the known background
solutions. At lowest order in perturbation one obtains linear equations for the perturbations, which are typically easier to analyse.
However, it is not guaranteed that all solutions to the linearized equations actually
arise as first approximations to solutions of the non-linear equations of the system.
Let us illustrate this point in a simple example~\cite{altas2018linearization}: For $(x,y) \in \mathbb{R}^2$, consider the algebraic system
$x(x^2 + y^2) = 0$, of which the exact solutions are $(0,y)$, where $y$ is any real number. On the other hand,
if we consider the linearized equations for perturbations $\delta x$ and $\delta y$ around a
background $(x_0, y_0)$, these satisfy $\delta x (x_0^2 + y_0^2) + x_0(2x_0 \delta x + 2y_0 \delta y) = 0$.
 If we let the background be $(x_0, y_0) = (0,0)$,
then any pairs $(\delta x, \delta y)$ satisfy
the linearized equation of motion, but those with $\delta x \neq 0$ cannot have arisen as linearizations of solutions to the
non-linear equation. One characterizes this phenomenon as the equation $x(x^2+y^2)=0$ being linearization unstable at $(0,0)$.

A well-known field-theoretic system with linearization instabilities is electrodynamics in a ``closed universe", i.e.\ in a spacetime with compact
Cauchy surfaces~\cite{brill1973instability,higuchi1991quantum1}.
Consider, for example, the electromagnetic field coupled to a charged matter field in such a spacetime.
Note that the conserved total charge of the matter field must vanish in this system. This fact is a simple consequence of
Gauss's law $\vec{\nabla}\cdot \vec{E} \propto \rho$, where $\vec{E}$ and  $\rho$ are the
electric field and charge density, respectively.  The integral
over a Cauchy surface of the charge density $\rho$
gives the total charge but the integral of $\vec{\nabla}\cdot\vec{E}$ vanishes because the Cauchy surface is compact.
At the level of the linearized theory about vanishing background electromagnetic and matter field the theory is non-interacting. Since the matter field is non-interacting,
there is no constraint on the total charge in the linearized theory, but a solution to the linearized (i.e.\ free) matter
field equation does not extend to
an exact solution to the full interacting theory unless its total charge $Q_{e}$
vanishes.  The linearization stability condition (LSC) in this case is $Q_{e}=0$.  If
a solution to the linearized equations satisfies this condition, then it extends to an exact solution to the full theory.
It is useful to note here that the charge $Q_{e}$ generates the global gauge symmetry of the free charged matter field.

In gravitational systems, it is known that such linearization instabilities occur
for any perturbations around a background which has both Killing symmetries and compact Cauchy
surfaces~\cite{brill1973instability, brill1972isolated,fischer1973linearization, moncrief1975spacetime, moncrief1976space, fischer1980structure, arms1981symmetry, arms1982structure, moncrief1978invariant, moncrief1979quantum}.
The LSCs which need to be imposed on such a closed universe are that the generators $Q$
of the background Killing symmetries must vanish when evaluated on any linearized solution.
For example, if the spacetime possesses time and space translation symmetries, then the corresponding conserved charges are
the energy and momentum, respectively.

In the late seventies Moncrief studied the r\^ole of these conditions in linearized \textit{quantum} gravity in spacetimes with compact Cauchy
surfaces.  He proposed that they should be imposed as physical-state conditions as in the Dirac
quantization~\cite{moncrief1978invariant, moncrief1979quantum}, i.e.
	\begin{equation}
		Q \ket{\mathrm{phys}} = 0\eqend{.}
	\end{equation}
Since the conserved charges $Q$ generate the spacetime symmetries (in the component of the identity),
he concluded that quantum linearization
stability conditions (QLSCs)
imply that all physical states must be invariant under the spacetime Killing symmetries.
He argued that these constraints can be viewed as a remnant of the diffeomorphism invariance of
the non-linear theory.

Imposing the invariance of the physical Hilbert space under the full background symmetries
(in the component of the identity) as required by the QLSCs
would appear too restrictive.
This problem is exemplified by de~Sitter space, a spacetime which is physically relevant for
inflationary cosmology.
This spacetime has Cauchy surfaces
with the topology of the $3$-dimensional sphere, which is compact.  Therefore,
the QLSCs
imply that all physical states of linearized quantum gravity on a
de~Sitter background ought to be invariant under the full $SO_0(4,1)$ symmetry group, i.e.\ the component
of the identity of $SO(4,1)$,
of the spacetime. However, this would appear to exclude all states except the
vacuum state, which would make the Hilbert space for the theory
quite empty~\cite{higuchi1991quantum1,moncrief1979quantum}.

However, there are non-trivial $SO_0(4,1)$ invariant states that have infinite norm and, hence, are not in the Hilbert
space.  Moncrief suggested that
a Hilbert space consisting of these invariant states could be constructed by dividing the infinite inner product by the infinite volume of the group $SO_0(4,1)$.
This suggestion was taken up in Ref.~\cite{higuchi1991quantum2}.  In that work a new inner product for $SO_0(4,1)$-invariant states was defined by
what  would later be termed group-averaging, which is an important ingredient in the refined algebraic quantization~\cite{ashtekar1995quantization} and
has been well studied in the context of Loop Quantum Gravity.
(The group-averaging procedure was also proposed in~\cite{lansman1993}.)
In this approach, one defines the invariant
states $\ket{\Psi}$ by starting with a non-invariant state $\ket{\psi}$ and averaging against the symmetry group $G$, assumed here to be a unimodular group
such as $SO_0(4,1)$,
to obtain
	\begin{equation}
		\ket{\Psi} = \int_G \mathrm{d}g \  U(g) \ket{\psi}\eqend{,}  \label{ga-state}
	\end{equation}
where $U$ is the unitary operator implementing the symmetry on the states.
The state $|\Psi\rangle$ can readily
be shown to be invariant, i.e.\ $\langle\phi| U(g)|\Psi\rangle = \langle\phi|\Psi\rangle$ for any state $|\phi\rangle$ in the
Hilbert space by the invariance of the measure $\mathrm{d}g$.
If the volume of the symmetry group is
finite, the inner product of the invariant states
$|\Psi_1\rangle$ and $|\Psi_2\rangle$ obtained from $|\psi_1\rangle$ and $|\psi_2\rangle$ as in \eqref{ga-state} is
	\begin{equation}
                     \langle\Psi_1|\Psi_2\rangle = V_G \int_G\mathrm{d}g\, \langle\psi_1|U(g)|\psi_2\rangle\eqend{,}
	\end{equation}
where $V_G$ is the volume of the group $G$.  If $G$ has infinite volume, e.g.\ if it is $SO_0(4,1)$, then
one needs to redefine the inner product on the invariant states by removing a factor of the group volume to
make them normalizable.  Thus, one defines the inner product on the new Hilbert space by
	\begin{equation}
		\braket{\Psi_1}{\Psi_2}_{ga} = \int_G \mathrm{d} g \ \bra{\psi_1} U(g) \ket{\psi_2}\eqend{.}
	\end{equation}
By this  group-averaging procedure one obtains an infinite-dimensional Hilbert space of $SO_0(4,1)$ invariant
states for linearized gravity in de~Sitter space~\cite{higuchi1991quantum1, higuchi1991quantum2}.
The group-averaging procedure was carried out for this spacetime also for other free
fields~\cite{marolf2009group1, marolf2009group2} to obtain Hilbert spaces
of invariant states. 
The QLSCs in de~Sitter space were also studied in the context of cosmological
perturbation~\cite{losicunruh2006,losic2007}.  The group-averaging procedure
has also been studied extensively in the context of constrained dynamical systems
(see. e.g.~\cite{louko2000,louko2005,louko2006,gomberoff1999}).
This method was extended to non-unimodular groups in~\cite{giulini-marolf1999}.

The group-averaging procedure can also be explicitly carried out for perturbative
quantum gravity in static space with topology of
$\mathbb{R} \times \mathrm{T}^3$, where the spatial Cauchy
surfaces are copies of $\mathrm{T}^3$, the $3$-dimensional torus,
to find the states satisfying the QLSCs.
The classical~\cite{brill1973instability} and quantum theory~\cite{moncrief1978invariant, higuchi1991linearized} of this
model have been studied and the group-averaging procedure can be carried out to obtain a physical Hilbert space of states invariant under the
$\mathbb{R} \times U(1)^3$ symmetry group of the
background.  Now, if one considers $4$-dimensional $\mathcal{N}=1$ simple
supergravity~\cite{deser1976SUGRA,freedmanetal1976SUGRA} on this background
spacetime, it is not difficult to see that
there are additional fermionic LSCs.  The purpose of this paper is to find all states satisfying the
bosonic and fermionic QLSCs and show that a Hilbert space of these states can be constructed using the
group-averaging procedure over the supergroup of symmetries of the linearized theory.

Let us describe how the LSCs arise for supergravity in this spacetime.
Recall that the energy and momentum of a system in general relativity can be expressed as an integral over a
$2$-dimensional surface at
infinity of a Cauchy surface (the ADM mass and momentum) in asymptotically-flat spacetime.  In classical perturbation theory
about Minkowski space,
this fact implies that the total energy and momentum of the linearized
fields can be expressed as a surface integral at infinity of perturbations of the next order.  Then, one expects that in
perturbation theory in the flat
$\mathbb{R}\times \mathrm{T}^3$ background, the total energy and momentum of the linearized fields vanish because
there is no spatial infinity.
This is indeed the case, and the vanishing of the total energy and momentum of the linearized field is expressed as the
LSCs.  Note here that the expression for the total energy, or the Hamiltonian,
in this spacetime is not positive definite and, therefore, can vanish for non-trivial field configurations.  Now,
in supergravity there is a spinor
supercharge $Q_\alpha$, $\alpha=1,2,3,4$, associated with a global supersymmetry variation, and it is known that this
supercharge can be written as an integral over $2$-dimensional surface at infinity of a Cauchy
surface in asymptotically-flat spacetime~\cite{hull1983the}.  This fact again implies that in static 3-torus space there are quadratic constraints on the linearized theory
corresponding to the vanishing of the supercharge, which are the fermionic LSCs.  In this paper we study
these fermionic LSCs together with the bosonic ones in linearized quantum supergravity.

The remainder of this paper is organized as follows.
 In section~\ref{constraint-derivation} we present a derivation of the (classical)
bosonic and fermionic LSCs for $\mathcal{N}=1$ simple supergravity in the background
of flat $\mathbb{R}\times \mathrm{T}^3$ spacetime.  We show that these conditions are of the form that the
conserved Noether charges of the linearized theory  
vanish.
In section~\ref{constraints} we discuss linearized supergravity in this spacetime and express the
LSCs in terms of the classical analogues of annihilation and creation operators.
In section~\ref{sec:bosonic-constraints} we impose the bosonic QLSCs on the states and recall how this can be understood in the context of group-averaging over the bosonic symmetry group.
In section~\ref{fermionic-constraints} we describe how all physcal states satisfying both the bosonic and fermionic
QLSCs are found and how a Hilbert space of physical states is constructed.
Then we show that the procedure of finding the physical Hilbert space can be interpreted as group-averaging over the
supergroup of global supersymmetry.  We summarize and discuss our results in section~\ref{summary}.
In \ref{App-A} we present a gauge transformation of the vierbein field that is proportional to the Lie derivative of a
vector field, though it is not necessary for this work.  In \ref{App-B1} we illustrate our derivation of the LSCs in the simple example
of electrodynamics in flat $\mathbb{R}\times \mathrm{T}^3$ spacetime. In \ref{App-B} we present the proof of
some identities used in this paper. In \ref{fermion-zero-mom-sector} we discuss some aspects of the zero-momentum sector of the gravitino field.
\ref{Appendix_C} presents an example of a two-particle state satisfying all QLSCs.
We follow the conventions of Ref.~\cite{freedman2012supergravity} throughout this paper.






\section{The linearization stability conditions} \label{constraint-derivation}

The action for the $4$-dimensional $\mathcal{N}=1$
simple supergravity~\cite{ freedman2012supergravity,van1981supergravity} with $8\pi G = 1$ is
	\begin{equation}
		S = \frac{1}{2} \int \mathrm{d}^4 x\,e\left[ R - \overline{\Psi}_\mu \gamma^{\mu \nu \rho} D_\nu \Psi_\rho
+ X_{(4\Psi)}
\right]\eqend{,} \label{action-SUGRA}
	\end{equation}
where $X_{(4\Psi)}$ consists of terms quartic in $\Psi_\mu$ (see e.g.~\cite{van1981supergravity} for the explicit form of $X_{(4\Psi)}$).
Here, $e^a_\mu$ are the vierbein fields, $e:= \textrm{det}(e_\mu^a)$, and
$\Psi_{\mu \alpha}$, $\alpha=1,2,3,4$, is the gravitino field, which is a Majorana spinor.
The
$4\times 4$ gamma matrices $\gamma^a$, $a=0,1,2,3$, satisfy the Clifford relation
$\{\gamma^a,\gamma^b\} = 2\eta^{ab}$, where
$\eta^{ab}=\textrm{diag}(-1,1,1,1)$.  The indices $a,b,c,\ldots$
are raised and lowered by the flat metric $\eta_{ab}$ whereas
the spacetime indices $\mu,\nu,\sigma,\ldots$ are raised and lowered by the spacetime metric
$g_{\mu\nu}= e_\mu^a e_{a \nu}$. The matrices $\gamma^a$ have the properties $\gamma^{0\dagger} = -\gamma^0$ and
$\gamma^{i\dagger} = \gamma^i$, $i=1,2,3$.  One defines $\gamma^\mu:= e^\mu_a \gamma^a$ and
$\gamma^{\mu\nu \rho} := \gamma^{[\mu}\gamma^\nu \gamma^{\rho]}$, where $[\cdots]$ indicates total
anti-symmetrization.  The Riemann tensor is given by
	\begin{equation}
		{R_{\mu\nu}}^{ab}:= \partial_\mu \omega_\nu^{\ ab} - \partial_\nu \omega_\mu^{\ ab}
+ \omega_\mu^{\ ac}\omega_{\nu c}^{\ \ b} - \omega_\nu^{\ ac}\omega_{\mu c}^{\ \ b}\eqend{,}
	\end{equation}
and the curvature scalar is $R := e^\mu_a e^\nu_b {R_{\mu\nu}}^{ab}$.  The spin connection
$\omega_\mu^{\ ab} = \omega_\mu^{\ [ab]}$ is expressed in terms of $e_\mu^a$ as
\begin{equation}
\omega_{\mu}^{\ ab} =  e^{a\nu}\partial_{[\mu}e_{\nu]}^{\ b} - e^{b\nu}\partial_{[\mu}e_{\nu]}^{\ a}
- \frac{1}{2}(e^{a\nu} e^{b\sigma} - e^{b\nu}e^{a\sigma})e_\mu^c \partial_\nu e_{c\sigma}\eqend{.}
\end{equation}
One also defines
$\overline{\Psi}  = \Psi^T C$, where $C$ is a unitary matrix satisfying $C^T = - C$ and $\gamma^{\mu T} = - C\gamma^\mu C^{-1}$.
The Majorana condition satisfied by $\Psi_\mu$ reads
$(\Psi_\mu^\dagger)_\alpha = \mathrm{i}(C\gamma^0\Psi_\mu)_\alpha$~\cite{van1981supergravity}.  We later choose
a Majorana representation for $\gamma^a$
in which $C = \mathrm{i}\gamma^0$.  In this representation we have $(\Psi_\mu^\dagger)_\alpha = (\Psi_\mu)_\alpha$.
The gravitino covariant derivative in \eqref{action-SUGRA} is given by
	\begin{equation}
		D_\mu \Psi_\nu = \partial_\mu \Psi_\nu + \frac{1}{4} \omega_{\mu a b} \gamma^{ab} \Psi_\nu\eqend{.}
\label{DPsi}
	\end{equation}
The action \eqref{action-SUGRA} is invariant under a local supersymmetry transformation of the form
	\begin{eqnarray}
		\Delta_\epsilon e_\mu^a & = & \frac{1}{2}\overline{\epsilon}\gamma^a \Psi_\mu\eqend{,} \label{super-1}\\
\Delta_\epsilon \Psi_\mu & = &  D_\mu \epsilon + Y_{\mu ab}\gamma^{ab}\epsilon\eqend{,}\label{super-2}
	\end{eqnarray}
where $Y_{\mu ab}$ is quadratic in $\Psi_\mu$ and
where $\epsilon$ is a spacetime dependent Grassmann variable with four components.
(See, e.g.~\cite{van1981supergravity} for the explicit form of $Y_{\mu ab}$.)

We consider this theory
on a purely bosonic static background whose spatial sections are flat $3$-dimensional tori.
For the background geometry we therefore take
	\begin{equation}
		\mathrm{d} s^2 = \eta_{\mu \nu} \mathrm{d} x^\mu \mathrm{d} x^\nu
= - \mathrm{d} t^2 + \mathrm{d}
x^2 + \mathrm{d} y^2 + \mathrm{d} z^2\eqend{,}
	\end{equation}
where the spatial coordinates $x$, $y$ and $z$ are periodic with periods $L_1, L_2$ and $L_3$ respectively, and we let
$V = L_1 L_2 L_3$ denote the spatial volume.

Let us first discuss the bosonic LSCs for this theory, which are well known as we described in section~\ref{intro}.
  The diffeomorphism transformation in the
direction of a vector $\zeta^\mu$ of the vierbein field $e_\mu^a$ and gravitino field $\Psi_\mu$
is\footnote{The transformation obtained as the commutator of two supersymmetry transformations is different from the diffeomorphism transformation
presented here (see, e.g.~\cite{freedman2012supergravity}).  The former takes the form
$\delta'_\zeta e_\mu^a = \delta_\zeta e_\mu^a - \zeta^\rho \omega_{\rho}^{\ ab}e_{b\,\mu}$,
$\delta'_\zeta \Psi_\mu = \delta_\zeta \Psi_\mu - \partial_\mu (\xi^\rho \Psi_\rho)$ to first order in the fields $\tilde{e}_\mu^a$ defined by
\eqref{vierbein-perturbed} and $\Psi_\mu$.  It can be shown that the LSCs corresponding to these two transformations are identical.}
\begin{eqnarray}
 \delta_\zeta e_\mu^a  & = & \zeta^\nu \partial_\nu e_\mu^a + e_\nu^a \partial_\mu \zeta^\nu\eqend{,} \label{diff-1} \\
\delta_\zeta \Psi_\mu  & = & \zeta^\nu \partial_\nu \Psi_\mu + \Psi_\nu \partial_\mu \zeta^\nu\eqend{.} \label{diff-2}
\end{eqnarray}
Now, we write the vierbein field as
\begin{equation}
e_\mu^a  =  \delta_\mu^a + \tilde{e}_\mu^{a}\eqend{.}  \label{vierbein-perturbed}
\end{equation}
That is, we write $e_\mu^a$ as the sum of its background value $\delta_\mu^a$ and perturbation $\tilde{e}_\mu^a$.  Then,
\begin{equation}
\delta_\zeta \tilde{e}_\mu^a = \delta_\nu^a \partial_\mu \zeta^\nu
+ \zeta^\nu \partial_\nu \tilde{e}_\mu^a + \tilde{e}_\nu^a \partial_\mu\zeta^\nu\eqend{.} \label{diff-e}
\end{equation}
It is important to note here that the part of $\delta_\zeta \tilde{e}_\mu^a$
that is independent of $\tilde{e}_\mu^a$ vanishes if $\zeta^\mu$ is a constant vector, i.e.\ a
Killing vector of the flat background.\footnote{If the background has non-zero curvature,
$\delta_\xi\tilde{e}_\mu^a$ does not necessarily vanish at lowest order even
if $\xi$ is a Killing vector of the background metric.
One needs to consider a transformation modified by an infinitesimal local Lorentz transformation in this case to make
$\delta_\xi \tilde{e}_\mu^a$ vanish at lowest order.  This is done in \ref{App-A}.}

Since the action (\ref{action-SUGRA}) is invariant under the diffeomorphism transformation given by \eqref{diff-2}
and \eqref{diff-e}, we have
\begin{equation}
\delta_\zeta \tilde{e}^a_\mu \frac{\delta S}{\delta \tilde{e}^a_\mu} + \delta_\zeta \Psi_\mu \frac{\delta S}{\delta\Psi_\mu}
=  \partial_\mu (\sqrt{-g}J_{(\zeta)}^\mu)\eqend{,} \label{identity}
\end{equation}
for some vector field $J_{(\zeta)}^\mu$, which is linear in $\zeta^\mu$.  Here the variation $\delta S/\delta\Psi_\mu$
is a left-variation, i.e.\
\begin{equation}
\delta S = \int \mathrm{d}^4x\, \delta\Psi_\mu \frac{\delta S}{\delta\Psi_\mu}\eqend{.}
\end{equation}
We adopt left-variations for fermionic fields throughout this paper.
We then expand $\delta_\zeta \tilde{e}_\mu^a$, $\delta_\zeta \Psi_\mu$,
$\delta S/\delta\tilde{e}_\mu^a$, $\delta S/\delta\Psi_\mu$ and $\sqrt{-g}J^\mu_{(\zeta)}$ according to the order in the fields
$\tilde{e}_\mu^a$ and $\Psi_\mu$, i.e.\ the number of these fields in the product, as
\begin{eqnarray}
&& \delta_\zeta \tilde{e}^a_\mu  =  \delta_\zeta^{(0)}\tilde{e}_\mu^a + \delta_\zeta^{(1)}\tilde{e}_\mu^a + \delta_\zeta^{(2)}\tilde{e}_\mu^a + \cdots
\eqend{,}\\
&& \delta_\zeta \Psi_\mu = \delta_\zeta^{(1)}\Psi_\mu + \delta_\zeta^{(2)}\Psi_\mu + \cdots\eqend{,}\\
&& \frac{\delta S}{\delta\tilde{e}_\mu^a}  =  E^{(1)\mu}_a + E^{(2)\mu}_a + \cdots\eqend{,}\\
&& \frac{\delta S}{\delta\Psi_\mu} = \mathcal{E}^{(1)\mu} + \mathcal{E}^{(2)\mu} + \cdots\eqend{,}\\
&& \sqrt{-g}\,J_{(\zeta)}^\mu  =  J_{(\zeta)}^{(0)\mu} + J_{(\zeta)}^{(1)\mu} + J_{(\zeta)}^{(2)\mu}+\cdots\eqend{,}
\end{eqnarray}
Recall that the background metric is flat.  We note that $E^{(0)\mu}_a = 0$ and
$\mathcal{E}^{(0)\mu} = 0$  because the flat spacetime with $\Psi_\mu = 0$ satisfies the field equations.  That is,
$\delta S/\delta \tilde{e}_\mu^a=0$ and $\delta S/\delta \Psi_\mu = 0$ if $\tilde{e}_\mu^a = 0$ and $\Psi_\mu=0$.

The identity \eqref{identity} must be satisfied order by order.  The first- and second-order equalities read
\begin{eqnarray}
&& (\delta_\zeta^{(0)}\tilde{e}_\mu^a)E_a^{(1)\mu} =  \partial_\mu J_{(\zeta)}^{(1)\mu}\eqend{,} \label{first-order-cons}\\
&& (\delta_\zeta^{(0)}\tilde{e}_\mu^a) E_a^{(2)\mu} + (\delta_\zeta^{(1)}\tilde{e}_\mu^a) E_a^{(1)\mu}
+ (\delta_\zeta^{(1)}\Psi_\mu) \mathcal{E}^{(1)\mu} =  \partial_\mu J_{(\zeta)}^{(2)\mu}\eqend{.}
\label{second-order-cons}
\end{eqnarray}
We emphasize here that these identities hold for any vector $\zeta^\mu$ and for any field configuration.
Now, if the vector field $\xi^\mu$ is a constant vector, then $\delta_\xi^{(0)}\tilde{e}_\mu^a = 0$ as can readily be seen from \eqref{diff-e}.
 Hence, by \eqref{first-order-cons}, the current $J_{(\xi)}^{(1)\mu}$ is conserved.
Then the charge
\begin{equation}
P_{(\xi)}^{(1)} := \int \mathrm{d}^3 \vec{x}\, J_{(\xi)}^{(1)0}\eqend{,}
\end{equation}
is conserved \textit{for any field configuration}.  In particular, it is conserved even if the fields are smoothly deformed to $0$ in the past or future of the $t=$ constant Cauchy surface where the integral is evaluated.  Since
$P_{(\xi)}^{(1)} = 0$ if the fields vanish, the conservation of this charge implies that
\begin{equation}
P_{(\xi)}^{(1)} = 0\eqend{,}  \label{1-conserved}
\end{equation}
\textit{for any field configuration} if $\xi^\mu$ is a constant vector.  The compactness of the Cauchy surfaces is crucial for this conclusion because this charge
is not necessarily conserved if the field configuration is time-dependent at infinity for the case where the Cauchy surfaces are non-compact.

Now, suppose one attempts to solve the field equations for $\tilde{e}_\mu^a$ and $\Psi_\mu$ order by order. Let
$(\tilde{e}_\mu^{(1)a}, \Psi_\mu^{(1)})$ be
a solution to the linearized equations and $(\tilde{e}_\mu^{(2)a},\Psi_\mu^{(2)})$ be the second-order correction.  Then,
\begin{eqnarray}
&& E_a^{(1)\mu}[\tilde{e}^{(1)}] = 0\eqend{,} \label{F-eq-1}\\
&& \mathcal{E}^{(1)\mu}[\Psi^{(1)}] = 0\eqend{,} \label{Fer-eq-1}\\
&& E_a^{(1)\mu}[\tilde{e}^{(2)}] + E_a^{(2)\mu}[\tilde{e}^{(1)},\Psi^{(1)}] = 0\eqend{,}  \label{F-eq-2}\\
&& \mathcal{E}^{(1)\mu}[\Psi^{(2)}] + \mathcal{E}^{(2)\mu}[\tilde{e}^{(1)},\Psi^{(1)}] = 0\eqend{.} \label{Fer-eq-2}
\end{eqnarray}
Here $\mathcal{E}^{(2)\mu}[\tilde{e}^{(1)},\Psi^{(1)}]$ is the vector-spinor $\mathcal{E}^{(2)\mu}$ evaluated with
$(\tilde{e}_\mu^a, \Psi_\mu) = (\tilde{e}_\mu^{(1)a},\Psi^{(1)}_\mu)$, and similarly for the others.
The identities~\eqref{first-order-cons} and \eqref{second-order-cons} imply
\begin{equation}
\partial_\mu ( J_{(\zeta)}^{(1)\mu}[\tilde{e}^{(2)}] + J_{(\zeta)}^{(2)\mu}[\tilde{e}^{(1)},\Psi^{(1)}]) = 0\eqend{,}
\end{equation}
\textit{for any vector field $\zeta^\mu$} as long as field equations \eqref{F-eq-1}-\eqref{F-eq-2} are satisfied.
 The charge corresponding to the conserved current
$J_{(\zeta)}^{(1)\mu}[\tilde{e}^{(2)}] + J_{(\zeta)}^{(2)\mu}[\tilde{e}^{(1)},\Psi^{(1)}]$ must vanish for any
$\zeta^\mu$ because this charge is
conserved even if the field $\zeta^\mu$ is smoothly deformed to zero in the past or future and is evaluated there.  Hence, if we define
\begin{equation}
P_{(\zeta)}^{(2)}[\tilde{e}^{(1)},\Psi^{(1)}] := \int\,\mathrm{d}^3\vec{x}\, J_{(\zeta)}^{(2)0}[\tilde{e}^{(1)},\Psi^{(1)}]\eqend{,}
\label{second-P}
\end{equation}
then
\begin{equation}
P_{(\zeta)}^{(2)}[\tilde{e}^{(1)},\Psi^{(1)}] = - P_{(\zeta)}^{(1)}[\tilde{e}^{(2)}]\eqend{,} \label{2-P-1}
\end{equation}
for any vector field $\zeta^\mu$ as long as the fields $(\tilde{e}_\nu^{(1)b},\Psi_\nu^{(1)})$ and $(\tilde{e}_\nu^{(2)b},\Psi_\nu^{(2)})$ are perturbative
solutions to the field equations.  In particular,  if $\zeta^\mu = \xi^\mu$ is a Killing vector, then
\begin{equation}
P_{(\xi)}^{(2)}[\tilde{e}^{(1)},\Psi^{(1)}] = 0\eqend{,} \label{2-conserved}
\end{equation}
because of \eqref{1-conserved}.  Equation~\eqref{2-conserved} is a consequence of the requirement that the solution
$(\tilde{e}_\nu^{(1)b},\Psi_\nu^{(1)})$
extend to an exact solution and does not follow from the linearized field equations.  This equation is called a linearization stability condition (LSC)
and has to be imposed on the solutions to the linearized equations.  In \ref{App-B1} we illustrate the argument leading to
\eqref{2-conserved} for electrodynamics.  The total charge of the linearized charged field must vanish in this model.

The LSCs from local supersymmetry can be derived in a similar manner.  We expand the supersymmetry transformation
given by \eqref{super-1} and \eqref{super-2} according to the number of fields in the product as
\begin{eqnarray}
\Delta_\epsilon \tilde{e}_\mu^a &  = & \Delta_\epsilon^{(1)}\tilde{e}_\mu^a + \Delta_\epsilon^{(2)} \tilde{e}_\mu^a + \cdots\eqend{,}\\
\Delta_\epsilon \Psi_\mu & = & \Delta_\epsilon^{(0)}\Psi_\mu + \Delta_\epsilon^{(1)}\Psi_\mu + \Delta_\epsilon^{(2)}\Psi_\mu + \cdots\eqend{.}
\end{eqnarray}
It is clear that $\Delta_\epsilon \tilde{e}_\mu^a = \Delta_\epsilon e_\mu^a$ given by \eqref{super-1} has no field independent contribution.  That is,
$\Delta_\epsilon^{(0)}\tilde{e}_\mu^a = 0$. The analogue of the identity~\eqref{identity} is
\begin{equation}
\Delta_\epsilon \tilde{e}_\mu^a \frac{\delta S}{\delta \tilde{e}_\mu^a} + \Delta_\epsilon \Psi_\mu \frac{\delta S}{\delta \Psi_\mu}
= \partial_\mu (\sqrt{-g}\mathcal{J}^\mu_{(\epsilon)})\eqend{.}
\end{equation}
From this equation one finds
\begin{eqnarray}
&& (\Delta_\epsilon^{(0)}\Psi_\mu)\mathcal{E}^{(1)\mu} = \partial_\mu \mathcal{J}_{(\epsilon)}^{(1)\mu}\eqend{,}\\
&& (\Delta_\epsilon^{(0)}\Psi_\mu)\mathcal{E}^{(2)\mu}
+ (\Delta_\epsilon^{(1)}\tilde{e}_\mu^a)E_a^{(1)\mu} + (\Delta_\epsilon^{(1)}\Psi_\mu)\mathcal{E}^{(1)\mu}
= \partial_\mu \mathcal{J}_{(\epsilon)}^{(2)\mu}\eqend{,} \label{second-order-consF}
\end{eqnarray}
for any spinor field $\epsilon$ and any field configuration. Since $\Delta_\varepsilon^{(0)}\Psi_\mu = \partial_\mu \varepsilon = 0$ if $\varepsilon$ is a
constant spinor, the charge defined by
\begin{equation}
Q_{(\varepsilon)}^{(1)} : = \int\,\mathrm{d}^3\vec{x}\, \mathcal{J}_{(\varepsilon)}^{(1)0}\eqend{,}
\end{equation}
vanishes for any field configuration if $\epsilon=\varepsilon$ is a constant spinor by the argument which led to
\eqref{1-conserved}.  By the same argument as that led to \eqref{2-P-1}, if we define
\begin{equation}
Q_{(\epsilon)}^{(2)}[\tilde{e}^{(1)},\Psi^{(1)}] := \int \mathrm{d}^3\vec{x}\,
\mathcal{J}_{(\epsilon)}^{(2)\mu}[\tilde{e}^{(1)},\Psi^{(1)}]\eqend{,}  \label{second-Q}
\end{equation}
then
\begin{equation}
Q_{(\epsilon)}^{(2)}[\tilde{e}^{(1)},\Psi^{(1)}] = - Q_{(\epsilon)}^{(1)}[\Psi^{(2)}]\eqend{,}
\end{equation}
for any spinor field $\epsilon$ if $(\tilde{e}_\nu^{(1)b},\Psi^{(1)}_\nu)$ gives a solution to the linearized field equations,
$E_a^{(1)\mu}[\tilde{e}^{(1)}] = \mathcal{E}^{(1)\mu}[\Psi^{(1)}] = 0$.
 In particular, if $\epsilon=\varepsilon$ is a constant spinor, since $Q_{(\varepsilon)}^{(1)}[\Psi^{(2)}] = 0$ in this case,
we must have
\begin{equation}
Q_{(\varepsilon)}^{(2)}[\tilde{e}^{(1)},\Psi^{(1)}] = 0\eqend{.}
\end{equation}
These are the LSCs arising from local supersymmetry on static $3$-torus space.

Next, we
shall derive the conserved currents $J_{(\xi)}^{(2)\mu}[\tilde{e}^{(1)},\Psi^{(1)}]$ given by
\eqref{second-order-cons} and
$\mathcal{J}_{(\varepsilon)}^{(2)\mu}[\tilde{e}^{(1)},\Psi^{(1)}]$ given by \eqref{second-order-consF} and the
corresponding conserved charges
$P^{(2)}_{(\xi)}[\tilde{e}^{(1)},\Psi^{(1)}]$ and $Q^{(2)}_{(\varepsilon)}[\tilde{e}^{(1)},\Psi^{(1)}]$
for a constant
vector $\xi^\mu$ and a constant spinor $\varepsilon$.  From now on we write
$\tilde{e}_{\mu}^{(1)a} = \tilde{e}_\mu^a$ and
$\Psi^{(1)}_\mu = \Psi_\mu$.

For either charge we only need the linearized field equations.  To find $E_a^{(1)\mu}$ for $\tilde{e}_\mu^a$, it is useful to
note that $eR$ in the Lagrangian
density in terms of the vierbein fields equals $\sqrt{-g}\,R$ given in terms of the metric tensor $g_{\mu\nu}$.  Thus,
we may vary $\sqrt{-g}\,R$ with respect to the metric tensor and then vary the metric tensor with respect to the vierbein fields.
Writing
$g_{\mu\nu} = \eta_{\mu\nu} + h_{\mu\nu}$, we find the perturbation $h_{\mu\nu}$ in terms of
$\tilde{e}_\mu^a$ at first order as
\begin{equation}
h_{\mu\nu} = \delta^a_\mu \tilde{e}_{a\nu} + \tilde{e}_{a\mu}\delta^a_\nu\eqend{.} \label{h-vary-by-e}
\end{equation}
Then we find
\begin{eqnarray}
E_a^{(1)\mu}\left[\tilde{e}\right] & = & \frac{1}{2}\delta^\nu_a
\left[ -\partial^\mu \partial^\lambda h_{\lambda\nu} - \partial_\nu\partial^\lambda h_{\lambda}^{\ \mu}
+ \partial^\mu\partial_\nu h + \Box h^{\mu}_{\ \nu} \right. \nonumber \\
&& \left. \ \ \ \ \ \ \ + \delta^\mu_\nu \partial^\lambda \partial^\sigma h_{\lambda\sigma} - \delta^\mu_\nu \Box h\right]\eqend{,} \label{e-equation}
\end{eqnarray}
where $h : = \eta^{\mu\nu}h_{\mu\nu}$ and $\Box : = \partial_\lambda\partial^\lambda$, with indices raised and lowered by $\eta_{\mu\nu}$.   We readily find
\begin{equation}
\mathcal{E}^{(1)\mu}\left[\Psi\right] = - C\gamma^{\mu\nu\rho}\partial_\nu \Psi_\rho\eqend{.} \label{Psi-equation}
\end{equation}

To find the bosonic conserved current $J^{(2)\mu}_{(\xi)}$,
we use \eqref{h-vary-by-e}-\eqref{Psi-equation} together with
\begin{eqnarray}
\delta_{(\xi)}^{(1)}\tilde{e}_\mu^a & = & \xi^\nu \partial_\nu \tilde{e}_\mu^a\eqend{,}\\
\delta_{(\xi)}^{(1)}\Psi_\mu & = & \xi^\nu \partial_\nu \Psi_\mu\eqend{,}
\end{eqnarray}
in \eqref{second-order-cons}, recalling $\delta^{(0)}_{(\xi)}\tilde{e}_\mu^a = 0$, to find
\begin{eqnarray}
\partial_\mu J_{(\xi)}^{(2)\mu}\left[\tilde{e},\Psi\right]
& = &  \frac{1}{4}(\xi^\lambda \partial_\lambda h_{\mu\nu})
\left[ -\partial^\mu \partial^\lambda h_{\lambda}^{\ \nu} - \partial^\nu\partial^\lambda h_{\lambda}^{\ \mu}
+ \partial^\mu\partial^\nu h + \Box h^{\mu\nu} \right. \nonumber \\
&& \left. \ \ \ \ \ \ \ \ \ \ + \eta^{\mu\nu} \partial^\lambda \partial^\sigma h_{\lambda\sigma} -
\eta^{\mu\nu} \Box h\right]
-(\xi^{\lambda}\partial_\lambda \overline{\Psi}_\mu) \gamma^{\mu\nu\rho}\partial_\nu \Psi_\rho\eqend{.}
\nonumber \\
\end{eqnarray}
The current $J_{(\xi)}^{(2)\mu}[h,\Psi]$, which depends on $\tilde{e}_\mu^a$ only through $h_{\mu\nu}$,
is identified with the Noether current for the spacetime translation in the direction of $\xi^\mu$ of the decoupled
theory consisting of a Fierz-Pauli Lagrangian for the graviton and a Rarita-Schwinger Lagrangian for the Majorana
gravitino~\cite{van1981supergravity}
	\begin{eqnarray}
		\mathcal{L} & = & - \frac{1}{4} \partial^\mu h \partial^\nu h_{\mu \nu} + \frac{1}{8} \partial_\mu h \partial^\mu h
+ \frac{1}{4} \partial_\sigma h^{\mu \sigma} \partial^\rho h_{\mu \rho} - \frac{1}{8} \partial_\rho h^{\mu \nu} \partial^\rho h_{\mu \nu}\nonumber \\
&&  - \frac{1}{2} \overline{\Psi}_\mu \gamma^{\mu \nu \rho} \partial_\nu \Psi_\rho\eqend{.}
\label{linear-Lag-SUSY}
	\end{eqnarray}
It can be given explicitly as
\begin{eqnarray} \fl \qquad 
J^{(2)\mu}_{(\xi)}\left[h,\Psi\right] & = & -(\xi^\lambda\partial_\lambda h_{\rho\sigma})\frac{\partial \mathcal{L}}{\partial(\partial_\mu h_{\rho\sigma})}
- (\xi^\lambda\partial_\lambda \Psi_\rho)\frac{\partial\mathcal{L}}{\partial(\partial_\mu \Psi_\rho)} + \xi^\mu \mathcal{L} \nonumber \\
& = & \frac{1}{4}\left[  \xi^\lambda\partial_\lambda h \partial_\nu h^{\mu\nu}
+ \xi^\lambda\partial_\lambda h^{\mu\rho}\partial_\rho h
- \xi^\lambda \partial_\lambda h \partial^\mu h \right.\nonumber \\
&& \left. \ \ \ - 2\xi^\lambda \partial_\lambda h^{\mu\rho} \partial^\sigma h_{\rho\sigma}
+ \xi^\lambda \partial_\lambda h_{\rho\sigma} \partial^\mu h^{\rho\sigma}\right] \nonumber \\
&& + \frac{1}{4}\xi^\mu\left[ - \partial^\rho h \partial^\sigma h_{\rho\sigma} + \frac{1}{2}\partial_\rho h\partial^\rho h
+ \partial_\sigma h^{\rho\sigma}\partial^\lambda h_{\rho\lambda} - \frac{1}{2}\partial_\lambda h_{\rho\sigma}\partial^\lambda h^{\rho\sigma}\right]
\nonumber \\
&& + \frac{1}{2}\left( \xi^\sigma \overline{\Psi}_\nu \gamma^{\nu\mu\rho}\partial_\sigma\Psi_\rho
- \xi^\mu \overline{\Psi}_\nu \gamma^{\nu\sigma\rho}\partial_\sigma \Psi_\rho\right)\eqend{.} \label{Jbose2}
\end{eqnarray}

Next we find the fermionic conserved current $\mathcal{J}_{(\varepsilon)}^{(2)\mu}$.
We note that the first-order part of the global supersymmetry transformation is given by
\begin{eqnarray}
\Delta^{(1)}_{(\varepsilon)}\tilde{e}_\mu^a & = & \frac{1}{2}\overline{\varepsilon}\gamma^a \Psi_\mu\eqend{,}\\
\Delta^{(1)}_{(\varepsilon)}\Psi_\mu & = & \frac{1}{4}\left(\partial_\mu \tilde{e}_{a\nu}\gamma^{\nu a}
+ \partial_\rho h_{\mu\nu}\gamma^{\nu\rho}\right)\varepsilon\eqend{.} \label{DeltaPsi1}
\end{eqnarray}
By substituting these formulas and equations \eqref{e-equation} and \eqref{Psi-equation} into \eqref{second-order-consF},
and using the identity~\cite{freedman2012supergravity}
\begin{equation}
\gamma^{\rho\nu}\gamma^{\sigma\mu\kappa}
= 3(\eta^{\nu[\sigma}\gamma^{\mu\kappa]\rho} - \eta^{\rho[\sigma}\gamma^{\mu\kappa]\nu}
+ \eta^{\nu[\sigma}\gamma^\mu \eta^{\kappa]\rho} - \eta^{\rho[\sigma}\gamma^\mu \eta^{\kappa]\nu})
\eqend{,}
\end{equation}
we find
\begin{equation}
\mathcal{J}^{(2)\mu}_{(\varepsilon)}[h,\Psi]
= \frac{1}{4}\left( \partial_\rho h_{\sigma\nu}\overline{\varepsilon}\gamma^{\rho\nu}
\gamma^{\mu\sigma\kappa}\Psi_\kappa  - \delta^a_\rho \tilde{e}_{a\nu}
\overline{\varepsilon}\gamma^{\rho\nu}\gamma^{\mu\sigma\kappa}\partial_\sigma \Psi_\kappa\right)\eqend{.}
\label{Jfermi2}
\end{equation}
Note that the second term vanishes if the (linear) local Lorentz invariance is fixed by requiring
$\tilde{e}_{a\nu} = \delta_a^\mu \delta_\nu^b \tilde{e}_{b\mu}$. (It vanishes by the linearized field equation for $\Psi_\mu$ as well.)
With this condition imposed,
the term which explicitly depends on $\tilde{e}_{b\nu}$ in the supersymmetry transformation
$\Delta^{(1)}_{(\varepsilon)}\Psi_\mu$ given by \eqref{DeltaPsi1} vanishes.  With this choice the current
$\mathcal{J}_{(\varepsilon)}^{(2)\mu}[h,\Psi]$ and the charge $Q_{(\varepsilon)}^{(2)}[h,\Psi]$
 are the Noether current and charge, respectively, for the supersymmetry transformation,
\begin{eqnarray}
\Delta^{(1)}_{(\varepsilon)}h_{\mu\nu} & = & \frac{1}{2}(\overline{\varepsilon}\gamma_\mu \Psi_\nu
+ \overline{\varepsilon}\gamma_\nu \Psi_\mu)\eqend{,}\\
\Delta^{(1)}_{(\varepsilon)}\Psi_\mu &  = & \frac{1}{4}\partial_\rho h_{\mu\nu}\gamma^{\nu\rho}\varepsilon
\eqend{,}
\end{eqnarray}
of the linearized supergravity Lagrangian given by \eqref{linear-Lag-SUSY}.

Thus, the classical LSCs are  $P_{(\xi)}^{(2)}[h,\Psi] = 0$ and
$Q_{(\varepsilon)}^{(2)}[h,\Psi] = 0$, where
$P_{(\xi)}^{(2)}[h,\Psi]$ and $Q_{(\varepsilon)}^{(2)}[h,\Psi]$ are the conserved charges corresponding to
the conserved currents $J_{(\xi)}^{(2)\mu}[h,\Psi]$ in \eqref{Jbose2} and
$\mathcal{J}_{(\varepsilon)}^{(2)\mu}[h,\Psi]$ in \eqref{Jfermi2} (without the second term)
respectively (see \eqref{second-P} and \eqref{second-Q}).  In the next section we discuss linearized
supergravity on static 3-torus space at the classical level
and express the LSCs in a form suitable for quantization.


\section{Linearized Supergravity on static $3$-torus space}\label{constraints}

The results of the previous section allow us to describe the LSCs entirely in terms of
the linearized theory.   By linearizing $\mathcal{N}=1$ simple supergravity in $4$ dimensions about
static $3$-torus background spacetime, we have the Lagrangian, which is given here again for convenience:
\begin{eqnarray}
		\mathcal{L} & = & - \frac{1}{4} \partial^\mu h \partial^\nu h_{\mu \nu} + \frac{1}{8} \partial_\mu h \partial^\mu h
+ \frac{1}{4} \partial_\sigma h^{\mu \sigma} \partial^\rho h_{\mu \rho} - \frac{1}{8} \partial_\rho h^{\mu \nu} \partial^\rho h_{\mu \nu}\nonumber \\
&&  - \frac{1}{2} \overline{\Psi}_\mu \gamma^{\mu \nu \rho} \partial_\nu \Psi_\rho\eqend{.}
\label{linear-Lagrangian}
	\end{eqnarray}
A suitable real Majorana representation for the $\gamma$-matrices is given by
	\begin{eqnarray}
		\gamma^0 = \left( \begin{array}{cc}
			0 & 1 \\
			-1 & 0
				\end{array} \right), \quad
		&&\gamma^1 = \left( \begin{array}{cc}
			1 & 0 \\
			0 & -1
				\end{array} \right), \nonumber \\
		\gamma^2 = \left( \begin{array}{cc}
			0 & \sigma^1 \\
			\sigma^1 & 0
				\end{array} \right), \quad
		&&\gamma^3 = \left( \begin{array}{cc}
			0 & \sigma^3 \\
			\sigma^3 & 0
				\end{array} \right)\eqend{,} \label{gamma-choice}
	\end{eqnarray}
where $\sigma^1$, $\sigma^2$ and $\sigma^3$ are the standard Pauli matrices.
In this representation the charge conjugation matrix takes the form
$C = \mathrm{i} \gamma^0$.
The action with the Lagrangian density \eqref{linear-Lagrangian} is invariant under the following gauge transformations:
	\begin{eqnarray}
		h_{\mu \nu} & \to & h_{\mu \nu} + \partial_\mu \zeta_\nu + \partial_\nu \zeta_\mu\eqend{,}\\
   \Psi_\mu & \to & \Psi_\mu + \partial_\mu \epsilon\eqend{,}
	\end{eqnarray}
where $\zeta^\mu$ is an arbitrary vector field and $\epsilon_\alpha$ is an arbitrary Majorana spinor field satisfying
$\epsilon^\dagger_\alpha = \epsilon_\alpha$.  This invariance
is a remnant of the diffeomorphism invariance and local supersymmetry of the full theory.  We shall fix this gauge
freedom completely.

Working on the spatial torus and imposing periodic boundary conditions on the fields allows us to decompose each field as a Fourier series. Due to the periodic boundary conditions, each field has a spatially constant zero-momentum component. The zero-momentum sector of the linearized theory will be seen to contain six bosonic and six fermionic degrees freedom and this sector forms an important ingredient in the QLSCs. Meanwhile, the non-zero momentum sector of the theory contains the usual gravitons and gravitinos, which each have two polarization states. Explicitly we expand the fields as
	\begin{eqnarray*}
		h_{\mu \nu}(t, \vec x) = \frac{1}{\sqrt{V}} h^{(0)}_{\mu \nu}(t) + \frac{1}{\sqrt{V}} \sum_{\vec k \neq 0} \tilde{h}_{\mu \nu}(t, \vec k) e^{\mathrm{i} \vec k \cdot \vec x}\eqend{,} \\
		\Psi_\mu(t, \vec x) = \frac{1}{\sqrt{V}} \psi_\mu(t) + \frac{1}{\sqrt{V}} \sum_{\vec k \neq 0} \tilde{\Psi}_{\mu}(t, \vec k) e^{\mathrm{i} \vec k \cdot \vec x}\eqend{,}
	\end{eqnarray*}	
	where the volume of the torus is $V = L_1 L_2 L_3$ and the $L_1$, $L_2$ and $L_3$ are the periods of the torus in the $x$-, $y$- and $z$-directions respectively.  The periodic boundary condition implies $\vec k = \left( \frac{2 \pi}{L_1} n_1, \frac{2 \pi}{L_2} n_2, \frac{2 \pi}{L_3} n_3 \right)$ with $n_1$, $n_2$ and $n_3$ integers, and reality restricts $h^{(0)}_{\mu \nu}$ to be real,
$\tilde{h}_{\mu \nu}(t, - \vec k)$ to be equal to $\tilde{h}_{\mu \nu}^\dagger (t, \vec k)$, and similarly for
$\psi_\mu$ and $\tilde{\Psi}_\mu(t,\vec{k})$.

The Lagrangian for the theory does not provide dynamical coupling between the modes with zero momentum and those with non-zero momentum $\vec k$.
Therefore, it is possible to analyse these separately. We begin with the zero-momentum sector of the theory.
We write $h_{\mu\nu}^{(0)}(t) = h_{\mu\nu}(t)$, dropping the
superscript ``$(0)$'', in the rest of this section.   The Lagrangian, i.e.\ the space integral of the Lagrangian density, for this sector reads
	\begin{equation}
		L_0 =  - \frac{1}{8} \partial_0 \tensor{h}{^i_i} \partial_0 \tensor{h}{^j_j} + \frac{1}{8} \partial_0 h^{ij} \partial_0
h_{ij} + \frac{1}{2} \bar{\psi}_i \gamma^{0} \gamma^{i j} \partial_0 \psi_j \eqend{.}
	\end{equation}
The classical~\cite{higuchi2001possible} and quantum~\cite{higuchi1991linearized} theory of the graviton contributions have been studied previously using the
ADM formalism without fixing the gauge.   Here we fix the gauge to extract the physical degrees of
freedom in the linearized theory. (This gauge fixing has little to do with the imposition of LSCs discussed later.)  

The linearized Lagrangian does not provide equations of motion for $h_{\mu 0}$ and $\psi_{0 \alpha}$, and these are precisely the components which carry the
gauge degrees of freedom. It is therefore possible to gauge-fix these components to vanish by solving $h_{00}(t)=2\partial_0 \xi_0(t)$,
$h_{i0}(t)=\partial_0 \xi_i(t)$, $i=1,2,3$ and
$\psi_0(t) = \partial_0\epsilon(t)$.  The elimination of $h_{0\mu}(t)$ and $\psi_0(t)$ exhausts the gauge freedom
in the zero-momentum sector, and all other components are physical. Thus,
in the zero-momentum sector the physical degrees of freedom are contained in the symmetric tensor $h_{i j}$ and
vector-spinor
$\psi_{i \alpha}$, $i=1,2,3$, each of which corresponds to six classical degrees of freedom.

The six bosonic degrees of freedom are contained in the symmetric tensor $h_{ij}$. To analyse this tensor it is
convenient to break it up into the trace and trace-free sectors.   Thus, we let
	\begin{equation}
		h_{ij}(t) = \sqrt{\frac{2}{3}} \delta_{ij}\,c(t) + 2 \sum_{A=1}^5 T^A_{ij} c_A(t)\eqend{,}
	\end{equation}
where the $T^A_{ij}$ are a set of five trace-free tensors which satisfy the orthonormality
condition $T^{A i j} T^B_{ i j} = \delta^{AB}$.  We choose them as
\begin{eqnarray}
T^1 = \frac{1}{\sqrt{2}}\left( \begin{array}{ccc} 1 & 0 & 0 \\ 0 & -1 & 0 \\ 0 & 0 & 0\end{array}\right),\
T^2 = \frac{1}{\sqrt{6}}\left(\begin{array}{ccc} 1 &  0 & 0 \\ 0 & 1 & 0 \\ 0 & 0 & -2\end{array}\right),\nonumber \\
T^3 = \frac{1}{\sqrt{2}}\left(\begin{array}{ccc}0 &  1  & 0 \\ 1 & 0 & 0 \\ 0 & 0 & 0\end{array}\right),\
T^4 = \frac{1}{\sqrt{2}}\left(\begin{array}{ccc} 0 & 0 & 1 \\ 0 & 0  & 0 \\ 1 & 0 & 0 \end{array}\right),\nonumber \\
T^5 = \frac{1}{\sqrt{2}}\left(\begin{array}{ccc} 0 & 0  & 0 \\ 0 & 0 & 1 \\ 0 & 1 & 0\end{array}\right).
\label{T-explicit}
\end{eqnarray}
In terms of the six variables $c$ and $c^A$, the Lagrangian reads
	\begin{equation}
		L_0 =  - \frac{1}{2} (\partial_0 c)^2 + \frac{1}{2}\sum_{A=1}^5 (\partial_0 c_A)^2 + \frac{1}{2} \bar{\psi}_i \gamma^0 \gamma^{ij} \partial_0 \psi_j \eqend{.}  \label{zero-mode-Lag}
	\end{equation}
The momentum conjugate to $h_{ij}$, i.e.\ $p^{ij} = \partial L_0/\partial(\partial_0 h_{ij})$, is
	\begin{eqnarray}
		p^{ij} & = &
 - \frac{1}{4} (\partial_0 \tensor{h}{^k_k})\delta^{ij}
+ \frac{1}{4} \partial_0 h^{ij}\nonumber \\
& = &  \frac{1}{2} \sqrt{\frac{2}{3}} \delta^{ij}\,c_P + \frac{1}{2} \sum_A T^{A ij}\, c_{PA}\eqend{,}  \label{h-zero-expansion}
	\end{eqnarray}
where $c_P = - \partial_0 c$ and $c_{PA} = \partial_0 c_A$ are the momenta conjugate to $c$ and $c_A$, respectively.
The time derivative of the zero-momentum field $h_{ij}$ is then
\begin{equation}
\partial_0 h_{ij} = 2\sum_{A=1}^5 T^{A}_{ij}\,c_{PA} - \sqrt{\frac{2}{3}}\delta_{ij}\,c_P\eqend{.} \label{zero-m-exp-h}
\end{equation}
The canonical Poisson bracket relations for $c$, $c_A$, $c_P$ and $c_{PA}$, and equivalently for $h_{ij}$ and $p^{kl}$, are
	\begin{eqnarray}
		\poisson{c}{c_P} = 1, \qquad \poisson{c_A}{c_{PB}} = \delta_{AB}\eqend{,} \label{cp-c}\\
		\poisson{h_{ij}}{p^{kl}} = \frac{1}{2} \left( \delta_{i}^{k} \delta_{j}^{ l} + \delta_{i}^{ l} \delta_{j}^{k}
\right)\eqend{.}
	\end{eqnarray}
	
The fermionic part of $L_0$ is of first order in time derivatives and therefore already defines a constrained dynamical system~\cite{dirac1964lectures, henneaux1994quantization}, and we need to use the Dirac bracket as the bracket to be ``promoted'' to the anti-commutator upon quantization.
The momentum variables $\pi^i_\alpha$ conjugate to $\psi_{i\alpha}$ are
	\begin{eqnarray}
		\pi^i_\alpha & = &\frac{\partial{L_0}}{\partial(\partial_0 \psi_{i \alpha})} \nonumber \\
&  = & - \frac{1}{2} (\bar{\psi}_j \gamma^0 \gamma^{ji})_\alpha\eqend{,}
	\end{eqnarray}
where the derivative of $L_0$ with respect to $\partial_0 \psi_{i\alpha}$ is the left derivative.
These momenta are not invertible in terms of the coordinates and velocities, so there are twelve primary constraints,
	\begin{equation}
		\phi^i_\alpha = \pi^i_\alpha + \frac{1}{2} (\bar{\psi}_j \gamma^0 \gamma^{ji})_\alpha \approx 0\eqend{.}	
	\end{equation}	
We note that the Poisson and Dirac brackets for two fermionic fields
are symmetric under the exchange of arguments.  That is, if the fields $A$ and $B$ are fermionic, then
$\{A,B\}_P = \{B,A\}_P$, and similarly for the Dirac bracket.
	
For the associated primary Hamiltonian $H_P$ we add arbitrary linear combinations of the primary constraints to the canonical Hamiltonian,
	\begin{equation}
		H_P = \partial_0 \psi_i \pi^i - L_0 + \lambda^i \phi_i = \lambda^i \phi_i\eqend{,}
	\end{equation}
where the $\lambda^i_\alpha$ are arbitrary. Notice in particular that the primary Hamiltonian for these modes
weakly vanishes, i.e.\ it vanishes if the constraints are satisfied.
The constraints are to be preserved under evolution by the primary Hamiltonian.  This requirement leads to consistency conditions
	\begin{equation}
		\partial_0 \phi^i \approx \poisson{\phi^i}{ H_P} \approx 0\eqend{.} \label{consistency-conditions}
	\end{equation}

To evaluate the Poisson bracket between the constraints, we use the canonical Poisson bracket,
	\begin{equation}
		\poisson{\psi_{i \alpha}}{\pi^j_\beta} = - \delta^i_j \delta_{\alpha \beta}\eqend{,}
	\end{equation}
which allows us to compute the Poisson bracket between the constraints (and therefore the primary Hamiltonian) as
	\begin{equation}
		\poisson{\phi^i_\alpha}{\phi^j_\beta} = - (C \gamma^0 \gamma^{ij})_{\alpha \beta}\eqend{.}
	\end{equation}
Hence, we find that the consistency conditions \eqref{consistency-conditions} imply $\lambda^i = 0$.  Thus, the primary Hamiltonian $H_P$ vanishes
and, as a result, the fields $\psi_i$ are time independent.  This fact can also be deduced from the Euler-Lagrange equation resulting from
\eqref{zero-mode-Lag}.
There are no secondary constraints and all the constraints are of second class. To obtain the bracket structure for the theory suitable for subsequent
quantization, we compute the Dirac bracket
	\begin{eqnarray}
		\dirac{\psi_{i \alpha}}{\pi^j_\beta} &= \poisson{\psi_{i \alpha}}{\pi^j_\beta} - \poisson{\psi_{i \alpha}}{\phi^k_\gamma} \left( \poisson{\phi}{\phi} \right)^{-1}_{k l \gamma \delta} \poisson{\phi^l_{\delta}}{\pi^j_\beta} \nonumber \\
		&= \frac{1}{2} \poisson{\psi_{i \alpha}}{\pi^j_\beta}\nonumber \\
&  = - \frac{1}{2} \delta^j_i \delta_{\alpha \beta}\eqend{.}
	\end{eqnarray}
On the Dirac bracket, we can impose the second-class constraints as strong conditions since the Dirac bracket between
second-class constraints vanishes.

Working explicitly in the Majorana representation, where $C = \mathrm{i} \gamma^0$, the Dirac bracket for $\psi_{i\alpha}$ evaluates to
	\begin{equation}
		\dirac{\psi_{i \alpha}}{\psi_{j \beta}} =
- \frac{\mathrm{i}}{2} \left( \delta_{ij} \delta_{\alpha \beta} - (\gamma_{i j})_{\alpha \beta} \right)\eqend{.}
	\end{equation}
To provide some insight into these relations we write
	\begin{equation}
		\psi_{i \alpha}
= \frac{1}{\sqrt{6}} (\gamma_i \eta)_\alpha + \sum_{A=1}^2 T_{ij}^A (\gamma^j \eta^A)_\alpha\eqend{,}
\label{zero-m-exp-psi}
	\end{equation}
where $\eta$ and $\eta_A$ are Majorana spinors.  The matrices $T^1$ and $T^2$ are given in  \eqref{T-explicit}.
These symmetric and traceless matrices satisfy
	\begin{equation}
		T^{Aij} \gamma_j T^B_{ik} \gamma^k = \delta^{AB}\eqend{.}
	\end{equation}
One can also show, by a component-by-component calculation,
\begin{equation}
\sum_{A=1}^2 T^A_{ik}\gamma^k T^A_{j\ell}\gamma^\ell = \frac{2}{3}\delta_{ij} - \frac{1}{3}\gamma_{ij}
\eqend{.}
\end{equation}
These relations can be used to show that the Dirac bracket relations for $\psi_{i\alpha}$ are equivalent to
	\begin{equation}\label{zero-mode-D-brackets}
		\dirac{\eta_\alpha}{\eta_{\beta}} =
\mathrm{i} \delta_{\alpha \beta}\eqend{,} \quad \dirac{\eta_\alpha^A}{\eta_\beta^B} = - \mathrm{i}
\delta^{AB}\delta_{\alpha\beta}\eqend{,}\quad \dirac{\eta_\alpha}{\eta_\beta^A} = 0\eqend{.} \label{dirac-relations}
	\end{equation}

Next, we briefly describe the mode expansion of the non-zero-momentum sector of this theory, which is well known. Letting the non-zero-momentum components of the fields be denoted by $\hat{h}_{\mu \nu}$ and $\hat{\Psi}_\mu$, we can completely fix the gauge freedom in this sector by imposing the following conditions
 (see, for instance,~\cite{freedman2012supergravity, van1981supergravity}):
	\begin{eqnarray}
		\hat{h}_{\mu 0} = \hat{h} = \partial^i \hat{h}_{ij} = 0\eqend{,}\\
 \gamma^i \hat{\Psi}_i = \hat{\Psi}_0 = \partial^i \hat{\Psi}_i = 0\eqend{.}
	\end{eqnarray}
Then we can write the expansion of the graviton 
as follows:
	\begin{eqnarray}
		\hat{h}_{ij}(\vec x, t) = \sqrt{\frac{2}{V}} \sum_{\vec k \neq 0} \sum_{\lambda = \pm} \frac{1}{\sqrt{k}}
\left[ H^\lambda_{ij}(\vec k) a_\lambda(\vec k) \mathrm{e}^{\mathrm{i} k \cdot x} + H^{\lambda *}_{ij}(\vec{k})
a_\lambda^\dagger(\vec k) \mathrm{e}^{-\mathrm{i} k \cdot x} \right]\eqend{,}
\label{h-expansion} 
	\end{eqnarray}
where $k:= |\vec k|$ and $k^\mu = (k, \vec k)$. We have let the complex conjugate of  $a_\lambda(\vec k)$ be denoted by $a^\dagger_\lambda(\vec k)$, anticipating quantization.
The symmetric and traceless polarization tensors are given by
\begin{equation}
H^\lambda_{ij}(\vec{k}) = \epsilon^\lambda_i (\vec k) \epsilon^\lambda_j(\vec k)\eqend{,}
\end{equation}
where the polarization vectors $\epsilon^{\lambda}_i(\vec k)$ are given by
\begin{equation}
\epsilon^{\pm}_i(\vec{k}) = \frac{1}{\sqrt{2}}(\hat{e}^{(1)}_i(\vec{k}) \pm \mathrm{i}\hat{e}^{(2)}_i(\vec{k}))\eqend{.}
\end{equation}
The unit spatial vectors $\hat{e}^{(1)}_i(\vec{k})$, $\hat{e}^{(2)}_i(\vec{k})$ and $\hat{e}^{(3)}_i(\vec{k}) = \vec{k}/k$
form a right-handed orthonormal system in this order.  The polarization vectors
$\epsilon^{\pm}_i(\vec{k})$ satisfy $\epsilon^{\lambda*}(\vec{k})\cdot \epsilon^{\lambda'}(\vec{k}) = \delta^{\lambda\lambda'}$ and
$k^i \epsilon^\lambda_i(\vec{k}) = 0$.
%
As a result,
$H^\lambda_{ij}(\vec{k})$ satisfy $k^i H^\lambda_{ij}(\vec k) = 0$ and
$H^{\lambda*}_{ij}(\vec k) H^{\lambda' ij}(\vec k) = \delta^{\lambda\lambda'}$.
The Lagrangian density for the non-zero-momentum sector of the graviton field in this gauge can be found from  \eqref{linear-Lagrangian} as
\begin{equation}
\mathcal{L}_{TT} = - \frac{1}{8}\partial_\rho \hat{h}^{ij}\partial^\rho \hat{h}_{ij}\eqend{.} \label{TT-Lagrangian}
\end{equation}
The standard procedure to find the Poisson bracket relations for the coefficients $a_\lambda(\vec{k})$ and $a^\dagger_\lambda(\vec{k})$ leads to
	\begin{equation}
		\poisson{a_\lambda(\vec k)}{a_\lambda^\dagger(\vec{k}^\prime)}
= - \mathrm{i} \delta_{\vec k, \vec{k}^\prime} \delta_{\lambda \lambda^\prime}
\eqend{,}  \label{graviton-poisson}
	\end{equation}
with all other brackets among $a_\lambda(\vec{k})$ and $a_\lambda^\dagger(\vec{k})$ vanishing.

The non-zero-momentum sector of the gravitino field can similarly be expanded into modes as
	\begin{eqnarray}
		\hat{\Psi}_{i \alpha} = \frac{1}{\sqrt{V}} \sum_{\vec k \neq 0} \sum_{\lambda = \pm}
\frac{1}{\sqrt{2k}} \left[ \epsilon_i^\lambda(\vec k) u^\lambda_\alpha (\vec k) b_\lambda(\vec k) \mathrm{e}^{\mathrm{i} k \cdot x}
+ \epsilon_i^{\lambda*}(\vec k) u^{\lambda*}_\alpha (\vec k) b_\lambda^\dagger(\vec k)
\mathrm{e}^{-\mathrm{i} k \cdot x} \right]\eqend{,} \nonumber \\ \label{Psi-expansion}
	\end{eqnarray}
where $u^\pm (\vec{k})$ are eigenspinors of $\gamma_5 = \mathrm{i}\gamma^0\gamma^1\gamma^2\gamma^3$ and
$\gamma^0 \hat{k}\cdot\vec{\gamma}$, where $\hat{k}:= \vec{k}/k$, with eigenvalues $\pm 1$ and $-1$, respectively.  We normalize them by requiring
$u^{\lambda\dagger}(\vec{k})u^{\lambda'}(\vec{k}') =2 k \delta^{\lambda\lambda'}$.
The fermionic coefficients $b_\lambda(\vec k)$ and $b^\dagger_\lambda(\vec k)$ satisfy the classical analogues of the anti-commutation relations for the
annihilation and creation operators:
	\begin{equation}
		\dirac{b_\lambda (\vec k)}{b_{\lambda^\prime}^\dagger(\vec{k}^\prime)} = -\mathrm{i} \delta_{\vec k, \vec{k}^\prime}
\delta_{\lambda \lambda^\prime}\eqend{,} \label{dirac-non-zero-mom}
	\end{equation}
with all other brackets among $b_\lambda(\vec{k})$ and $b_\lambda^{\dagger}(\vec{k})$ vanishing.

\section{Imposing the bosonic linearization stability conditions} \label{sec:bosonic-constraints}

We quantize the linearized theory by promoting the physical degrees of freedom to operators acting on some Hilbert space, and we impose
	\begin{equation}
		[\textrm{(anti)-commutator}]_{\pm} = \mathrm{i} \hbar \{ \textrm{Poisson/Dirac Bracket} \}\eqend{,}
	\end{equation}
as the algebraic relations between the operators, where $[A,B]_{\pm}:= AB \pm BA$. We work in units with $\hbar = 1$.
 Thus, equations \eqref{cp-c} and \ref{graviton-poisson} become
\begin{eqnarray}
\left[ c, c_P\right]_{-} = \mathrm{i},\ \ \ \left[ c_A, c_{PB}\right]_{-} = \mathrm{i}\delta_{AB}\eqend{,}\\
\left[ a_\lambda(\vec{k}), a^\dagger_{\lambda'}(\vec{k}')\right]_{-} = \delta_{\lambda\lambda'}\delta_{\vec{k},\vec{k}'}\eqend{,}
\end{eqnarray}
respectively. On the other hand, the relations \eqref{dirac-relations} and \eqref{dirac-non-zero-mom} become
\begin{eqnarray}
\left[\eta_\alpha,\eta_\beta\right]_+ = - \delta_{\alpha\beta},\ \ \left[ \eta_\alpha^A,\eta^B_\beta\right]_+ = \delta^{AB}\delta_{\alpha\beta},\ \
\left[ \eta_\alpha,\eta^A_\beta\right]_+ =0\eqend{,}  \label{fermi0-anti-com}\\
\left[ b_\lambda(\vec{k}),b_{\lambda'}^\dagger(\vec{k}')\right]_{+} = \delta_{\lambda\lambda'}\delta_{\vec{k},\vec{k}'}\eqend{,}
\end{eqnarray}
respectively.

After imposing our gauge conditions and using the field equations, the time component of the conserved
Noether current $J_{(\xi)}^{(2)\mu}[h,\Psi]$ for the spacetime translation symmetry becomes
\begin{eqnarray}
J_{(\xi)}^{(2)0} & = & \frac{\sqrt{V}}{8}\xi_0\left[
\partial_0 h_{ij}\partial_0 h^{ij} -  \partial_0\tensor{h}{^j_j}\partial_0 \tensor{h}{^i_i}\right] \nonumber \\
&&
- \frac{1}{4}\xi_\nu\partial^\nu \hat{h}^{ij}\partial_0 \hat{h}_{ij} + \frac{1}{8}\xi_0 \partial_\nu \hat{h}_{ij}\partial^\nu \hat{h}^{ij}
- \frac{\mathrm{i}}{2}\xi_\nu \hat{\Psi}^T_i\partial^\nu \hat{\Psi}^i\eqend{,} \label{J2-explicit}
\end{eqnarray}
where we have dropped the cross terms between the zero-momentum and non-zero-momentum sectors because they do not contribute to the space
integral of $J_{(\xi)}^{(2)0}$, which gives the conserved charge.
Let us write
\begin{equation}
\xi_0 H + \xi_i P^i = \int\mathrm{d}^3\vec{x}\, J^{(2)0}_{(\xi)}\eqend{.}
\end{equation}
By substituting \eqref{h-zero-expansion}, \eqref{h-expansion} and \eqref{Psi-expansion} into \eqref{J2-explicit} and integrating the result over space,
one finds
	\begin{eqnarray}
		H &= - \frac{1}{2} c_P^2 + \sum_{A = 1}^5 \frac{1}{2} c_{P A}^2 + \sum_{\vec k \neq 0} \sum_{\lambda = \pm} |\vec k |
\left( a^\dagger_\lambda(\vec k) a_\lambda(\vec k) + b^\dagger_\lambda(\vec k) b_\lambda(\vec k) \right)\eqend{,} \label{Hamiltonian-constraint} \\
		\vec P &= \sum_{\vec k \neq 0} \sum_{\lambda = \pm} \vec k \left( a^\dagger_\lambda(\vec k) a_\lambda(\vec k)
+ b^\dagger_\lambda(\vec k) b_\lambda(\vec k) \right)\eqend{,} \label{momentum-constraint}
	\end{eqnarray}
the bosonic sector of which agrees with the conserved charges given in~\cite{higuchi1991linearized}.
Note that the zero-point energy which had to be renormalized away in the pure-gravity case is absent here because of supersymmetry.  Note also
that the Hamiltonian $H$ is not positive definite.

The Hilbert space for the theory can be constructed as the tensor product of a non-zero momentum sector and a sector with zero
momentum. In the sector with non-zero momentum,
we have the usual Fock spaces for the gravitons and gravitinos with the number operators for the gravitons and gravitinos
(with a given momentum and a polarization) being
$a_\lambda^\dagger(\vec{k})a_\lambda(\vec{k})$ and $b_\lambda^\dagger(\vec{k})b_\lambda(\vec{k})$, respectively.
A suitable basis of states is
given by states with definite number of particles in each momentum and polarization. Thus, the normalized state with $n_{B\lambda}(\vec{k})$ gravitons and
$n_{F\lambda}(\vec{k})$ gravitinos with momentum $\vec{k}$  and polarization $\lambda$ can be given as
	\begin{equation}
		\ket{ \{ n_{B} \} } \otimes \ket{ \{ n_{F} \} } = \prod_{\vec{k},\lambda=\pm}
\left[\frac{1}{\sqrt{n_{B\lambda}(\vec{k})!}}(a_\lambda^{\dagger}(\vec{k}))^{n_{B\lambda}(\vec{k})}
 (b_\lambda^\dagger(\vec{k}))^{n_{F\lambda}(\vec{k})}\right]
|0\rangle\eqend{,}
	\end{equation}
where the vacuum state $|0\rangle$ satisfies $a_\lambda(\vec{k})|0\rangle = b_\lambda(\vec{k})|0\rangle = 0$ for all $\vec{k}$ and $\lambda$.
(For the gravitino $n_{F\lambda}(\vec{k})= 0$ or $1$, of course.)

For the graviton zero-modes, we represent the commutation rules
$\com{c}{c_P} = \mathrm{i}$ and $\com{c_A}{c_{P B}} = \mathrm{i} \delta_{AB}$
on wave functions which are functions of the variables $c$ and $c_A$ by
	\begin{eqnarray}
		c \ \mapsto \textrm{multiply by } c, \qquad c_P  \mapsto - \mathrm{i} \fpartial{}{c}, \\
		c_A \mapsto \textrm{multiply by } c_A, \quad c_{PA} \mapsto -\mathrm{i} \fpartial{}{c_A}\eqend{.}		
	\end{eqnarray}
Therefore, if we take some state which is proportional to a single eigenstate of the number operators,
	\begin{equation}
		\ket{\textrm{state}} = \Psi \otimes \ket{ \{ n_{B} \} } \otimes \ket{ \{ n_{F} \} }\eqend{,}  \label{general-states}
	\end{equation}		
where $\Psi$ is some wave function of $c$ and $c_A$, then the Hamiltonian and momentum operators,
\eqref{Hamiltonian-constraint} and \eqref{momentum-constraint}, acting on such a state take the form
	\begin{eqnarray}
		H \ket{\textrm{state}} = \frac{1}{2} \left( + \frac{\partial^2}{\partial c^2}
- \sum_{A = 1}^{5} \frac{\partial^2}{\partial c_A^2} + M^2 \right) \ket{\textrm{state}}\eqend{,} \\
		\vec P \ket{\textrm{state}} = \sum_{\vec k \neq 0} \sum_{\lambda = \pm} \vec k \left( n_{B \lambda}(\vec k) + n_{F \lambda}(\vec k) \right)
\ket{\textrm{state}}\eqend{,}
	\end{eqnarray}
where we have defined a ``squared mass" for these states by
	\begin{equation}
	M^2 = 2\sum_{\vec k \neq 0} \sum_{\lambda = \pm} |\vec k| \left( n_{B \lambda}(\vec k) + n_{F \lambda}(\vec k) \right)\eqend{.}
\label{mass-squared}
	\end{equation}
The operators $H$ and $\vec{P}$ do not contain the zero-momentum sector of the gravitino field.  They appear in the global supercharges as we shall see
in the next section.

As discussed in section \ref{constraint-derivation}, the conserved charges $H$ and $\vec{P}$
must vanish classically for the classical solutions to the linearized equations if they were to be
extendible to exact solutions. Moncrief proposed that in quantum theory these linearization stability conditions (LSCs)
should be imposed as constraints on the
physical states.  Thus, the constraints on the physical states $|\textrm{phys}\rangle$ are
\begin{equation}
		H \ket{\textrm{phys}} = 0, \quad \vec P \ket{\textrm{phys}} = 0\eqend{.} \label{bosonic-constraints}
	\end{equation}
In~\cite{higuchi1991linearized} the group-averaging procedure was used to find all states satisfying these constraints and define an inner product among these
states for pure gravity on static $3$-torus space.  We apply this procedure to $\mathcal{N}=1$ simple supergravity in
this section.

We start with the Hilbert space $\mathcal{H}_0$ of superpositions of the states defined by \eqref{general-states}.  Take two states in this Hilbert space
of the form $|\varphi_1\rangle = \Psi_1(c,c_A) \otimes |\Phi_1\rangle$ and $|\varphi_2\rangle = \Psi_2(c,c_A)\otimes |\Phi_2\rangle$, where $|\Phi_1\rangle$
and $|\Phi_2\rangle$ are states in the Fock space $\mathcal{F}$ of non-zero-momentum gravitons and gravitinos.  The inner product between these states is
\begin{equation}
\langle \varphi_1|\varphi_2\rangle_{\mathcal{H}_0} = \langle\Phi_1|\Phi_2\rangle_{\mathcal{F}} \int \mathrm{d}c\mathrm{d}^5\vec{c}\,
\Psi^*_1\Psi_2\eqend{,}
\end{equation}
where $\vec{c}$ is the vector with components $c_A$, $A=1,2,3,4,5$.
Let us choose both $|\Phi_1\rangle$ and $|\Phi_2\rangle$ to have a definite value of $M^2$ defined by \eqref{mass-squared}.
(If these states have different values of $M^2$, then $\langle \Phi_1|\Phi_2\rangle_{\mathcal{F}} = 0$.)  Then,
by expressing $\Psi_I(c,\vec{c})$, $I=1,2$, as a Fourier integral in the six-dimensional space with coordinates $(c,\vec{c})$, these states can be given as
\begin{equation}
|\varphi_I\rangle = \int \frac{\mathrm{d}p^0}{2\pi}\frac{\mathrm{d}^5\vec{p}}{(2\pi)^5}
F_I(p^0,\vec{p})\mathrm{e}^{-\mathrm{i}p^0 c + \mathrm{i}\vec{p}\cdot \vec{c}}\otimes |\Phi_I\rangle\eqend{,} \label{Fourier}
\end{equation}
where $\vec{p}$ is also a $5$-dimensional vector and $\vec{p}\cdot\vec{c} = p^A c_A$.  The inner product for these states is
\begin{equation}
\langle\varphi_1|\varphi_2\rangle_{\mathcal{H}_0} =  \langle \Phi_1|\Phi_2\rangle_{\mathcal{F}}
\int \frac{\mathrm{d}p^0}{2\pi}\frac{\mathrm{d}^5\vec{p}}{(2\pi)^5} F_1^*(p^0,\vec{p})F_2(p^0,\vec{p})\eqend{.} \label{original-norm}
\end{equation}

Since the operators $H$ and $\vec{P}$ are the generators of spacetime translations, we can construct states satisfying \eqref{bosonic-constraints} by
averaging the states $|\varphi_I\rangle$ over this translation group as follows:
\begin{eqnarray}
|\varphi_I^{(B)}\rangle = \frac{1}{2V}\int_{-\infty}^\infty \mathrm{d}\alpha^0
\left(\prod_{i=1}^3 \int_0^{L_i}\mathrm{d}\alpha_i\right)\exp(\mathrm{i}\alpha^0 H - \mathrm{i}\vec{\alpha}\cdot\vec{P})
|\varphi_I\rangle\eqend{.}
\end{eqnarray}
The integral over space acts as the projector onto the sector of the Fock space with zero total momentum:
\begin{equation}
\frac{1}{V}\left(\prod_{i=1}^3 \int_0^{L_i}\mathrm{d}\alpha_i\right)\exp(- \mathrm{i}\vec{\alpha}\cdot\vec{P})|\Phi_I\rangle
= \mathcal{P}_{\vec{P}=0}|\Phi_I\rangle\eqend{.}
\end{equation}
The integral over $\alpha^0$ can readily be evaluated using the representation \eqref{Fourier} since
\begin{equation}
H \mathrm{e}^{-\mathrm{i}p^0 c + \mathrm{i}\vec{p}\cdot\vec{c}}\otimes |\Phi_I\rangle =
\frac{1}{2}\left[ - (p^0)^2 + \vec{p}^2 + M^2\right] \mathrm{e}^{-\mathrm{i}p^0 c + \mathrm{i}\vec{p}\cdot\vec{c}}\otimes |\Phi_I\rangle\eqend{.}
\end{equation}
Thus we find, with the notation $|\Phi_I^{(\vec{P}=0)}\rangle = \mathcal{P}_{\vec{P}=0}|\Phi_I\rangle$,
\begin{eqnarray}
|\varphi_I^{(B)}\rangle & = &  \int \frac{\mathrm{d}p^0}{2\pi}\frac{\mathrm{d}^5\vec{p}}{(2\pi)^5}
2\pi \delta((p^0)^2 - \vec{p}^2 - M^2) F_I(p^0,\vec{p})\mathrm{e}^{-\mathrm{i}p^0 c + \mathrm{i}\vec{p}\cdot \vec{c}}\otimes
|\Phi_I^{(\vec{P}=0)}\rangle \nonumber
\\
& = &  \int \frac{\mathrm{d}^5\vec{p}}{(2\pi)^5}
\left[ f_I^{(+)}(\vec{p})
\mathrm{e}^{-\mathrm{i}E(\vec{p}) c + \mathrm{i}\vec{p}\cdot \vec{c}}
+ f_I^{(-)}(\vec{p})
\mathrm{e}^{\mathrm{i}E(\vec{p})c + \mathrm{i}\vec{p}\cdot\vec{c}}\right]\otimes |\Phi_I^{(\vec{P}=0)}\rangle \eqend{,} \nonumber \\
\label{invariant-state}
\end{eqnarray}
where $E(\vec{p}) = \sqrt{\vec{p}^2 + M^2}$ and where
\begin{equation}
f_I^{(\pm)}(\vec{p}) = \frac{F_I(\pm E(\vec{p}),\vec{p})}{2E(\vec{p})}\eqend{.}
\end{equation}
Although we obtained the invariant state $|\varphi^{(B)}_I\rangle$ by averaging $|\varphi_I\rangle$
over the spacetime translation group, it is easy to show that any state with
definite value of $M^2$ satisfying the constraints \eqref{bosonic-constraints} is of this form. (The zero-momentum sector is a solution to the
$6$-dimensional Klein-Gordon equation with mass $M$.)

The states $|\varphi^{(B)}_I\rangle$ are indeed invariant, i.e.\ they satisfy the QLSCs
given by \eqref{bosonic-constraints}.  However, they have infinite norm and, hence,
are not in the Hilbert space $\mathcal{H}_0$: the inner product
$\langle\varphi^{(B)}_I|\varphi^{(B)}_I\rangle_{\mathcal{H}_0}$ computed using
\eqref{original-norm} is infinite  because of the $\delta$-function in \eqref{invariant-state}.  The group-averaging inner product for the Hilbert space
$\mathcal{H}_B$ of invariant states is defined by
\begin{eqnarray} \fl \qquad 
\langle \varphi_1^{(B)}|\varphi_2^{(B)}\rangle_{\mathcal{H}_B}
& =  &  \frac{1}{2V}\int_{-\infty}^\infty \mathrm{d}\alpha^0
\left(\prod_{i=1}^3 \int_0^{L_i}\mathrm{d}\alpha_i\right)
\langle \varphi_1|\exp(\mathrm{i}\alpha^0 H - \mathrm{i}\vec{\alpha}\cdot\vec{P})|\varphi_2\rangle_{\mathcal{H}_0}\nonumber \\
& = & \langle \varphi_1|\varphi^{(B)}_2\rangle_{\mathcal{H}_0} \nonumber \\
& = &  2\int \frac{\mathrm{d}^5\vec{p}}{(2\pi)^5}E(\vec{p})
\left[ f_1^{(+)*}(\vec{p})f_2^{(+)}(\vec{p}) + f_1^{(-)*}(\vec{p})f_2^{(-)}(\vec{p})\right] \nonumber \\
&& \times
\langle \Phi_1^{(\vec{P}=0)}|\Phi_2^{(\vec{P}=0)}\rangle_{\mathcal{F}}\eqend{.} \label{GA-inner-product}
\end{eqnarray}
Notice that, although this inner product is defined in terms of the ``seed states'' $|\varphi_I\rangle$, it depends only on the invariant states
$|\varphi^{(B)}_I\rangle$.  This inner product is equivalent to that proposed in~\cite{marolf1995cosmology} in the context of quantum cosmology.

Since the constraint $H|\textrm{phys}\rangle=0$ is a Klein-Gordon equation, it is tempting to use the Klein-Gordon inner product for the Hilbert space
$\mathcal{H}_B$:
\begin{equation}
\langle \varphi^{(B)}_1|\varphi^{(B)}_2\rangle_{KG}
= \mathrm{i}\int \mathrm{d}^5 c
\left( \Psi_1^*\frac{\partial\Psi_2}{\partial c} - \frac{\partial\Psi_1^*}{\partial c}\Psi_2\right)
\langle\Phi_1^{(\vec{P}=0)}|\Psi_2^{(\vec{P}=0)}\rangle_{\mathcal{F}}\eqend{.}
\end{equation}
This inner product can readily be evaluated as
\begin{eqnarray}
\langle \varphi_1^{(B)}|\varphi_2^{(B)}\rangle_{KG}
& =  &  2\int \frac{\mathrm{d}^5\vec{p}}{(2\pi)^5}E(\vec{p})
\left[ f_1^{(+)*}(\vec{p})f_2^{(+)}(\vec{p}) - f_1^{(-)*}(\vec{p})f_2^{(-)}(\vec{p})\right] \nonumber \\
&& \times
\langle \Phi_1^{(\vec{P}=0)}|\Phi_2^{(\vec{P}=0)}\rangle_{\mathcal{F}}\eqend{,}
\end{eqnarray}
which is identical with the group-averaging inner product, $\langle \varphi_1^{(B)}|\varphi_2^{(B)}\rangle_{\mathcal{H}_B}$, given by
\eqref{GA-inner-product} except for the minus sign in the second term.  Although the Klein-Gordon inner product is more widely used in quantum
cosmology, it can be used only if the space of geometries considered has the structure of spacetime~\cite{kuchar1991}.
(In our example, it is the $6$-dimensional
Minkowski space.)
The group-averaging inner product, which we use in this paper,  has wider applicability.  For example, it can be used in
recollapsing quantum cosmology as shown in~\cite{marolf1995cosmology}.

In the next section we impose the fermionic QLSCs derived in the previous section.  We shall find that the states satisfying these constraints and
the inner product among them can be found by group-averaging over the relevant supergroup.

\section{Imposing the fermionic linearization stability conditions} \label{fermionic-constraints}

The time component of the conserved fermionic current $\mathcal{J}_{(\varepsilon)}^{(2)\mu}[h,\Psi]$ is
\begin{eqnarray}
\mathcal{J}_{(\varepsilon)}^{(2)0}\left[h,\Psi\right]
&  = & \frac{1}{4}\left[ \partial_0 h\overline{\varepsilon}\gamma^i \psi_i - \partial_0 h_{ij}\overline{\varepsilon} \gamma^i \psi^j
- \partial_0 \hat{h}_{ij}\overline{\varepsilon}\gamma^j \hat{\Psi}^i + \partial_\ell \hat{h}_{ij}\overline{\varepsilon}\gamma^0 \gamma^\ell \gamma^j
\hat{\Psi}^i\right] \eqend{,}\nonumber \\
\label{super-current}
\end{eqnarray}
where we dropped the cross terms between the zero-momentum and non-zero-momentum sectors since they do not contribute to the space integral.
By letting
\begin{equation}
\overline{\varepsilon}Q = - \int\,\mathrm{d}^3\vec{x}\mathcal{J}_{(\varepsilon)}^{(2)0}\eqend{,}
\end{equation}
and substituting \eqref{zero-m-exp-h} and \eqref{zero-m-exp-psi} into
\eqref{super-current} and integrating over space, we find
\begin{equation}
		Q = Q^{(0)}
+ \widehat{Q}\eqend{,}
\end{equation}
with the zero- and non-zero-momentum contributions respectively given by
\begin{eqnarray}
Q^{(0)} &= \frac{1}{2} \left( c_P \eta + \sum_{A = 1}^5 \sum_{B=1}^2 \tensor{T}{^{Ai}_j} T^B_{ik} \gamma^j \gamma^k c_{PA} \eta^B \right) \eqend{,} \label{eq:Q0-explicit}\\
\widehat{Q} &= \frac{1}{4}\int \mathrm{d}^3\vec{x}\,
\left( \partial_0\hat{h}^{ij}\gamma_j \hat{\Psi}_i - \hat{h}^{ij} \gamma_j \partial_0 \hat{\Psi}_i\right)\eqend{,}
\end{eqnarray}
where we have integrated by parts in the second term of $\widehat{Q}$ and used $\gamma^j \partial_j \hat{\Psi}_i = - \gamma^0 \partial_0 \hat{\Psi}_i$.

One is readily able to find the mode expansion for $\widehat{Q}$, which then yields
\begin{eqnarray}
Q & =  \frac{1}{2} \left( c_p \eta + \sum_{A = 1}^5 \sum_{B=1}^2 \tensor{T}{^{Ai}_j} T^B_{ik} \gamma^j \gamma^k c_{PA} \eta^B \right)
\nonumber \\
		&\quad -  \frac{\mathrm{i}}{2}
\sum_{\vec k \neq 0} \sum_{\lambda = \pm} \left[ \epsilon^\lambda(\vec k)\cdot\vec{\gamma} u^{\lambda*}(\vec k) a_\lambda (\vec{k})
b_{\lambda}^\dagger(\vec k) - \epsilon^{\lambda *}(\vec k)\cdot\vec{\gamma} u^\lambda(\vec k) a_\lambda^\dagger(\vec k) b_\lambda (\vec k) \right]
\eqend{.} \nonumber \\
\label{explicit-Q}
	\end{eqnarray}
This charge and the bosonic charges, $P^\mu = (H, \vec{P})$, satisfy the supersymmetry algebra.  That is,
$[P^\mu, Q]_{-} = 0$, $[P^\mu,P^\nu]_{-}=0$ and
	\begin{equation}
		\left[ Q_\alpha,Q_\beta\right]_{+} = \frac{\mathrm{1}}{2} (\gamma_\mu\gamma^0)_{\alpha \beta} P^\mu\eqend{.} \label{SUSY-algebra}
	\end{equation}
Since the contribution to $Q$ with different $\vec{k}$ anti-commute, equation~\eqref{SUSY-algebra} holds for each $\vec{k}$ including $\vec{k}=0$.
In deriving \eqref{SUSY-algebra} we have used the following identities for $\vec{k}=0$ and $\vec{k}\neq 0$ proved in \ref{App-B}:
\begin{eqnarray}
\sum_{B=1}^2 \sum_{A=1}^5\sum_{A'=1}^5 T\indices{^A^i_j}T^B_{ik} T^{A^\prime}_{i' j'} T\indices{^{B}^{i'}_{k'}} \gamma^j\gamma^k \gamma^{k'}\gamma^{j'}
c_{PA}c_{PA'}  = \sum_{A=1}^5 (c_{PA})^2\eqend{,}  \label{interesting-identity} \\
\left[\epsilon^{\pm*}(\vec{k})\cdot \vec{\gamma} u^\pm(\vec{k})\right]_\alpha
\left[\epsilon^{\pm}(\vec{k})\cdot\vec{\gamma} u^{\pm*}(\vec{k})\right]_\beta
+ (\alpha \leftrightarrow \beta)  = 2(\gamma\cdot k \gamma^0)_{\alpha\beta}\eqend{.}
\label{standard-identity}
\end{eqnarray}
We also recall that $\gamma^\mu$ are real, $\gamma_5$ is purely imaginary and
that $\epsilon^{\pm*}(\vec{k}) = \epsilon^{\mp}(\vec{k})$ and $u^{\pm*}(\vec{k}) = u^{\mp}(\vec{k})$.

One can represent the anti-commutation relations \eqref{fermi0-anti-com} satisfied by 
the twelve fermionic operators $\eta_\alpha$ and $\eta^A_\alpha$ in the zero-momentum sector of the gravitino field
by a $64$-dimensional indefinite-metric Hilbert space (or Krein space) $\mathcal{H}_{0F}$ 
as shown in \ref{fermion-zero-mom-sector}.  
The basis vectors of this space
can be chosen such that $32$ of them are normalized with postive norm and the other $32$ are normalized with
negative norm. [We say here that $|v\rangle$ is normalized with positive (negative) norm if
$\langle v|v\rangle = 1$ ($\langle v|v\rangle = -1$).]   An important property of $\mathcal{H}_{0F}$ used later is that
the $16$-dimensional subspace of $\mathcal{H}_{0F}$ of states annihilated by the operators $d_1$ and $d_2$ defined by
\begin{equation}\label{eq:def-c}
d_1 := (\eta_1 + \mathrm{i}\eta_2)/\sqrt{2}\eqend{,}\  \  d_2=(\eta_3 + \mathrm{i}\eta_4)/\sqrt{2}\eqend{,}
\end{equation}
is a positive-norm subspace
because this subspace is spanned by the states of the form \eqref{eq:general-state} with $n_1=n_2=0$ (with $M=2$)
[see \eqref{eq:general-norm}].
The Hilbert space of states satifying the bosonic QLSCs is in fact
the tensor product $\mathcal{H}_B\otimes\mathcal{H}_{0F}$.  We call this tensor product space also
$\mathcal{H}_B$ from now on in order not to complicate the notation.

Now we are in a position to find the physical states $|\textrm{phys}\rangle$ satisfying the fermionic QLSCs,
$Q_\alpha|\textrm{phys}\rangle = 0$, as well as the bosonic ones, $P^\mu|\textrm{phys}\rangle = 0$.   Since the operators $P^\mu$ commute
with $Q_\alpha$, we specialize to the Hilbert space $\mathcal{H}_B$ of states satisfying the bosonic QLSCs and
identify $P^\mu$ with the null operator. 
Thus, we have $\left[ Q_\alpha,Q_\beta\right]_{+}=0$ in place of \eqref{SUSY-algebra}.
Splitting the supercharge into zero- and non-zero-momentum contributions, these anti-commutation relations of the supercharge can be rewritten as
	\begin{equation}
		[Q^{(0)}_\alpha, Q^{(0)}_\beta]_{+} = - \frac{1}{4} M^2 \delta_{\alpha \beta} = - [\widehat{Q}_\alpha, \widehat{Q}_\beta]_{+}\eqend{,} 
	\end{equation}
where $M^2$ is defined by \eqref{mass-squared}.  We specialize to an eigenspace of the operator $M^2$ without loss of
generality.  Thus, we treat $M$ as if it were a non-negative number.

Assume $M > 0$.  Then, in each common eigenspace of the operators $c_P$ and $c_{PA}$ with eigenvalues
$p_0$ and $p_{A}$, respectively, satisfying $p_0^2 - \vec{p}^2 =M^2$,
the operators $Q^{(0)}_\alpha$ given by \eqref{eq:Q0-explicit} is of the form
\begin{equation}
Q^{(0)}_\alpha = \frac{M}{2}\left( \pm \eta_\alpha \cosh\theta+ \sum_{B=1}^2 \sum_{\beta=1}^4
{C_\alpha}^\beta_B \eta_\beta^B\sinh\theta\right)\eqend{,}
\end{equation}
where ${C_\alpha}^\beta_B$ can be regarded as a $4\times 8$ matrix with the rows and columns 
labeled by $\alpha$
and $(B,\beta)$, respectively.  Here, $\cosh\theta = |c_P|/M$ and $\sinh\theta = |\vec{c}_P|/M$, where
$\vec{c}_P = (c_{P1},c_{P2},c_{P3},c_{P4},c_{P5})$.  
The four $8$-dimensional column vectors of the matrix ${C_\alpha}^\beta_B$ are orthonormal, i.e.\ 
\begin{equation}
\sum_{B=1}^2 \sum_{\beta=1}^4 {C_{\alpha_1}}^\beta_B {C_{\alpha_2}}^\beta_B = \delta_{\alpha_1\alpha_2}
\eqend{.}
\end{equation}
Then by \ref{fermion-zero-mom-sector} there is a unitary operator $U$, i.e.\ an operator preserving the inner product,
on $\mathcal{H}_{0F}$ such that 
\begin{equation} \label{eq:unitary-relation}
Q^{(0)}_\alpha = (M/2)U\eta_\alpha U^\dagger\eqend{.}
\end{equation}

Instead of directly working with $Q^{(0)}_\alpha$ and $\widehat{Q}_\alpha$, it is more convenient to combine them into annihilation- and creation-type operators~\cite{henneaux1994quantization} as
	\begin{eqnarray}
		a_1 &= \frac{\sqrt{2}}{M} (Q^{(0)}_1 +\mathrm{i} Q^{(0)}_2), \qquad  & a_2 = \frac{\sqrt{2}}{M} \left( Q^{(0)}_3 + \mathrm{i} Q^{(0)}_4 \right)\eqend{,} \label{eq:a_i-def}\\
		b_1 &= \frac{\sqrt{2}}{M} ( \widehat{Q}_1 +\mathrm{i} \widehat{Q}_2), \qquad  & b_2 = \frac{\sqrt{2}}{M} ( \widehat{Q}_3 +\mathrm{i} \widehat{Q}_4)\eqend{,}
	\end{eqnarray}
provided that $M > 0$.
\footnote{For $M = 0$, formally one can proceed in a similar manner by considering ($\delta$-function normalisable) plane wave states $\Psi(c, \vec c) = e^{\pm \mathrm{i} p c + \mathrm{i} \vec p \cdot \vec c}$. On such states the supercharge takes the form $Q = \frac{p}{2} ( \pm \eta + R)$, with $[\eta_\alpha,\eta_\beta]_{+} = - \delta_{\alpha\beta}$ and 
$\anticom{R_\alpha}{R_\beta} = \delta_{\alpha \beta}$.   Then, we define ladder-type operators
$a_1 = (\eta_1+\mathrm{i}\eta_2)/\sqrt{2}$,
$a_2 = (\eta_3+\mathrm{i}\eta_4)/\sqrt{2}$, $b_1 = (R_1 + \mathrm{i}R_2)/\sqrt{2}$ and
$b_2 = (R_3 + \mathrm{i} R_4)/\sqrt{2}$ and proceed in a manner similar to the case with $M>0$.}
%
These new operators satisfy the anti-commutation relations of Fermi oscillators (up to a sign), i.e.\
$a_i^2=b_i^2 = a_i^{\dagger 2} = b_i^{\dagger 2} = 0$ and
	\begin{equation}
		[a_i,a_j^\dagger]_{+} = - \delta_{ij}, \qquad [b_i, b_j^\dagger]_{+} = \delta_{ij}\eqend{.}
	\end{equation}

Now we construct  all states $|\varphi^{(BF)}\rangle\in \mathcal{H}_{B}$ satisfying $Q_\alpha|\varphi^{(BF)}\rangle = 0$, $\alpha=1,2,3,4$.
These constraints can be organized as
	\begin{equation}
		(a_i + b_i) \ket{\varphi^{(BF)}} = (a_i^\dagger + b_i^\dagger) \ket{\varphi^{(BF)}} = 0, \quad i = 1,2\eqend{.} \label{ab-constraints}
	\end{equation}
Note that all states are linear combinations of the states each annihilated by either $b_1$ or $b_1^\dagger$ since
$|\varphi^{(B)}\rangle =  b_1^\dagger b_1|\varphi^{(B)}\rangle + b_1 b_1^\dagger|\varphi^{(B)}\rangle$ for any state
$|\varphi^{(B)}\rangle\in\mathcal{H}_{B}$.
The same is true for each pair of operators, $(b_2,b_2^\dagger)$, $(a_1,a_1^\dagger)$ and $(a_2,a_2^\dagger)$.  This means that a general state
is a superposition of states, each belonging to a $16$-dimensional Fock space
built on a state $|\chi^{(B)}\rangle$ satisfying $a_i|\chi^{(B)}\rangle = b_i|\chi^{(B)}\rangle = 0$, $i=1,2$,
 by applying the creation-type operators $a_i^\dagger$ and $b_i^\dagger$.   
Notice that equations \eqref{eq:unitary-relation} and \eqref{eq:a_i-def} imply
$a_1 = Ud_1 U^\dagger$ and
$a_2 = Ud_2 U^\dagger$, where $U$ is a unitary operator on $\mathcal{H}_{0F}$ and where
$d_1$ and $d_2$ are defined by \eqref{eq:def-c}.
Since the states annihilated by $d_1$ and $d_2$ have positive norm as
we stated before, we have
$\langle\chi^{(B)}|\chi^{(B)}\rangle_{\mathcal{H}_{B}} > 0$.

We label the $16$ possible states in such a Fock space as
follows:
\begin{equation}( a_1^\dagger)^{m_1}(a_2^\dagger)^{m_2}(b_1^\dagger)^{ n_1}(b_2^{\dagger})^{n_2}|\chi^{(B)}\rangle
= |m_1m_2n_1n_2\rangle\eqend{,} \label{eq:Fock-basis}
\end{equation}
with each $m_1, m_2, n_1, n_2$ being either $0$ or $1$. For example, we define
	\begin{equation}
		\ket{\chi^{(B)}} = \ket{0000}, \ \ a_1^\dagger \ket{\chi^{(B)}} = \ket{1000},
 \ \ a_1^\dagger a_2^\dagger b_1^\dagger b_2^\dagger \ket{\chi^{(B)}} =
\ket{1111}\eqend{.}
	\end{equation}
Thus, we look for states satisfying the fermionic QLSCs in the form
\begin{equation}
|\varphi^{(BF)}\rangle = \sum_{m_1=0}^1\sum_{m_2=0}^1 \sum_{n_1=0}^1\sum_{n_2=0}^1 c_{m_1m_2n_1n_2}|m_1m_2n_1n_2\rangle\eqend{.}
\end{equation}
We find
\begin{eqnarray}
(a_1+b_1)|\varphi^{(BF)}\rangle & =   \sum_{m_1=0}^1\sum_{m_2=0}^1 \sum_{n_1=0}^1\sum_{n_2=0}^1
c_{m_1m_2n_1n_2} \nonumber \\
&  \ \ \ \   \times \left[- m_1|0m_2n_1n_2\rangle
+ (-1)^{m_1+m_2}n_1|m_1m_2 0 n_2\rangle\right]\eqend{.}
\end{eqnarray}
The coefficient of the term $|0m_21n_2\rangle$ in this equation
is $-c_{1m_21n_2}$.  Hence $c_{1m1n} = 0$ for all $m$ and $n$.  We can conclude similarly that
$c_{0m0n} = c_{m0n0} = c_{m1n1} = 0$ by using the other constraints.  Thus,  $c_{m_1m_2n_1n_2} = 0$ if $m_1=n_1$ or $m_2=n_2$.  Hence
the invariant state $|\varphi^{(BF)}\rangle \in \mathcal{H}_{B}$, i.e.\ the state satisfying the fermionic QLSCs, must be of the following form:
	\begin{equation}
		\ket{\varphi^{(BF)}} = A \ket{1100} + B \ket{0110} + C \ket{1001} + D \ket{0011}\eqend{,}
\label{gen-inv-fermi}
	\end{equation}
where $A$, $B$, $C$ and $D$ are constants.
Note that the four states $|1100\rangle$, $|0110\rangle$, $|1001\rangle$ and $|0011\rangle$ are mutually orthogonal
and satisfy
\begin{equation}
\eqalign{\langle 1100|1100\rangle = \langle 0011|0011\rangle &
= \langle\chi^{(B)}|\chi^{(B)}\rangle_{\mathcal{H}_B} \eqend{,} \\  \langle 0110|0110\rangle
= \langle 1001|1001\rangle &= - \langle\chi^{(B)}|\chi^{(B)}\rangle_{\mathcal{H}_B}
\eqend{.}}
\end{equation}
In particular, the two states $|0110\rangle$ and $|1001\rangle$ have negative norm.

By applying the constraints \eqref{ab-constraints} we find the following unique solution up to an overall normalization:
	\begin{equation}
		\ket{\varphi^{(BF)}}\propto |\varphi^{(BF)}_{P}\rangle : = \ket{1100} - \ket{0110} + \ket{1001} + \ket{0011}\eqend{.} \label{varphiP}
	\end{equation}
Thus, 
there is precisely one possible combination which satisfies the fermionic QLSCs in each Fock space spanned by $(a_1^\dagger)^{n_1}(a_2^\dagger)^{n_2}(b_1^\dagger)^{n_1}(b_2^\dagger)^{n_2}|\chi^{(B)}\rangle$, where $|\chi^{(B)}\rangle$ is any
state with fixed positive $M^2$, satisfying the bosonic QLSCs and the conditions
\begin{eqnarray}
	(Q^{(0)}_1 + \mathrm{i} Q^{(0)}_2)|\chi^{(B)}\rangle 
= (Q^{(0)}_3 + \mathrm{i} Q^{(0)}_4)|\chi^{(B)}\rangle = 0\eqend{,}\\
(\widehat{Q}_1 + \mathrm{i} \widehat{Q}_2) |\chi^{(B)}\rangle 
= (\widehat{Q}_1 + \mathrm{i} \widehat{Q}_2)|\chi^{(B)}\rangle = 0\eqend{.}
\end{eqnarray}

Now, in section~\ref{sec:bosonic-constraints} it was shown that all states satisfying the bosonic QLSCs are obtained by group-averaging.
Here we show that the same is true for the fermionic QLSCs.  That is, the state $|\varphi^{(BF)}\rangle$ in \eqref{gen-inv-fermi} is obtained as
\begin{equation}
|\varphi^{(BF)}\rangle  =  - \int \mathrm{d}^4\theta\, \mathrm{e}^{- \overline{\theta}Q}|\varphi^{(B)}\rangle\eqend{,} \label{minus-sign}
\end{equation}
for some linear combination $|\varphi^{(B)}\rangle$ of the states $|m_1m_2n_1n_2\rangle$,
where $\theta=(\theta_1,\theta_2,\theta_3,\theta_4)$ are Grassmann numbers and where
$\overline{\theta} = \mathrm{i}\theta^T\gamma^0$ and
$\mathrm{d}^4\theta = \mathrm{d}\theta_4\mathrm{d}\theta_3\mathrm{d}\theta_2\mathrm{d}\theta_1$.
(The minus sign in \eqref{minus-sign} has been introduced for convenience.)
By the usual rules, $\int\mathrm{d}\theta_\alpha = 0$ and $\int \mathrm{d}\theta_\alpha \theta_\alpha = 1$, $\alpha=1,2,3,4$, this equation becomes
\begin{eqnarray}
|\varphi^{(BF)}\rangle & = & - Q_1Q_2Q_3Q_4|\varphi^{(B)}\rangle \nonumber \\
& = &  \frac{1}{4}(Q_1- \mathrm{i}Q_2)(Q_1 +\mathrm{i} Q_2)(Q_3- \mathrm{i}Q_4)(Q_3 + \mathrm{i}Q_4)|\varphi^{(B)}\rangle\nonumber \\
& = &  \frac{M^4}{16}(a_1^\dagger + b_1^\dagger)(a_1 + b_1)(a_2^\dagger + b_2^\dagger)(a_2 + b_2)|\varphi^{(B)}\rangle\eqend{.}
\end{eqnarray}
One readily finds that $Q_1Q_2Q_3Q_4|m_1m_2n_1n_2\rangle = 0$ unless
$|m_1m_2n_1n_2\rangle = |1100\rangle$, $|0110\rangle$, $|1001\rangle$ or $|0011\rangle$ and that
	\begin{eqnarray}
		(a_1^\dagger + b_1^\dagger)(a_1 + b_1)(a_2^\dagger + b_2^\dagger)(a_2 + b_2) \ket{1100} &
= |\varphi_{P}^{(BF)}\rangle\eqend{,} \label{1-e}\\
		(a_1^\dagger + b_1^\dagger)(a_1 + b_1)(a_2^\dagger + b_2^\dagger)(a_2 + b_2) \ket{0110} &= |\varphi_{P}^{(BF)}\rangle\eqend{,} \\
		(a_1^\dagger + b_1^\dagger)(a_1 + b_1)(a_2^\dagger + b_2^\dagger)(a_2 + b_2) \ket{1001}
&= - |\varphi_{P}^{(BF)}\rangle\eqend{,}	\\
		(a_1^\dagger + b_1^\dagger)(a_1 + b_1)(a_2^\dagger + b_2^\dagger)(a_2 + b_2) \ket{0011} &= |\varphi_{P}^{(BF)}\rangle\eqend{,}
\label{4-e}
	\end{eqnarray}
where the state $|\varphi_{P}^{(BF)}\rangle$ is defined by \eqref{varphiP}.
Thus, starting from any linear combination $|\varphi^{(B)}\rangle$ of $|m_1m_2n_1n_2\rangle$, we find
	\begin{equation}
		- \int \mathrm{d}^4\theta\,\mathrm{e}^{-\overline{\theta}Q} \ket{\varphi^{(B)}} =
- Q_1Q_2Q_3Q_4\ket{\varphi^{(B)}} = \kappa |\varphi_{P}^{(BF)}\rangle\eqend{,}	\label{superG}
	\end{equation}
with $\kappa \in \mathbb{C}$.

Now, an explicit computation shows that $\langle \varphi_{P}^{(BF)}|\varphi_{P}^{(BF)}\rangle_{\mathcal{H}_{B}} = 0$.
In fact this can be deduced immediately from
\eqref{superG} because $Q_\alpha^2 = 0$ on the space $\mathcal{H}_B$ of states satisfying the bosonic QLSCs.  However, the group-averaging
formula \eqref{superG}
suggests that one can proceed in analogy with the bosonic case to define a new inner product.  Thus, for two states
$|\varphi^{(B)}_1\rangle, |\varphi_2^{(B)}\rangle \in \mathcal{H}_{B}$ we obtain two states satisfying the fermionic QLSCs as
\begin{equation}
|\varphi_I^{(BF)}\rangle =  - Q_1Q_2Q_3Q_4|\varphi_I^{(B)}\rangle\eqend{,}
\end{equation}
and define the new inner product by
\begin{eqnarray}
\langle \varphi_1^{(BF)}|\varphi_2^{(BF)}\rangle_{\mathcal{H}_{BF}} & =
& - \langle \varphi^{(B)}_1|Q_1Q_2Q_3Q_4|\varphi_2^{(B)}\rangle_{\mathcal{H}_B} \nonumber \\
& = & \langle \varphi_1^{(B)}|\varphi_2^{(BF)}\rangle_{\mathcal{H}_{B}}\eqend{.}
\end{eqnarray}
Equations \eqref{1-e}-\eqref{4-e} and \eqref{varphiP} imply that, if
\begin{equation}
|\varphi^{(B)}_I\rangle = \kappa^{(1)}_I|1100\rangle +\kappa^{(2)}_I|0110\rangle
+ \kappa^{(3)}_I|1001\rangle + \kappa^{(4)}_I|0011\rangle\eqend{,}
\end{equation}
then
\begin{equation}
|\varphi_I^{(BF)}\rangle =\frac{M^4}{16}\lambda_I|\varphi_{P}^{(B)}\rangle\eqend{,}
\end{equation}
where
\begin{equation}
\lambda_I = \kappa_I^{(1)} + \kappa_I^{(2)} - \kappa_I^{(3)} + \kappa_I^{(4)}\eqend{,} \label{kappa-sum}
\end{equation}
 and
\begin{equation}
\langle \varphi_1^{(BF)}|\varphi_2^{(BF)}\rangle_{\mathcal{H}_{BF}}
=\frac{M^4}{16} \lambda_1^* \lambda_2\langle\chi^{(B)} |\chi^{(B)}\rangle\eqend{.}  \label{inner-explicit}
\end{equation}
Thus, our new inner product $\langle\cdot|\cdot\rangle_{\mathcal{H}_{BF}}$ is
positive definite.\footnote{The definition of a positive-definite inner product used here is similar to that  in Appendix D of Ref.~\cite{gibbonshiguchi2015}}
Although it is defined
in terms of the ``seed states'' $|\varphi_I^{(B)}\rangle$, which do not satisfy the fermionic QLSCs,
the new inner product depends only on the invariant states $|\varphi_I^{(BF)}\rangle$.

We constructed the states satisfying the bosonic and fermionic QLSCs step by step.
We first found the states satisfying the bosonic QLSCs and an inner product among them in the previous section, following~\cite{higuchi1991linearized}.
Then, we found the states satisfying the fermionic QLSCs as well among these states and defined an inner product
for these invariant states.  Our construction can in fact be understood as group-averaging over the supergroup generated by
$P^\mu$ and $Q_\alpha$.
Starting from a state  $\ket{\varphi_I}$, $I=1,2$, in the original Hilbert space $\mathcal{H}_0$,
we define a state satisfying all constraints
by integrating over the supergroup~\cite{fayet1977supersymmetry, rogers2007supermanifolds}:
	\begin{equation}
		|\varphi_I^{(BF)}\rangle = - \frac{1}{2V}\int \mathrm{d}^4 \alpha \ \mathrm{d}^4 \theta
\exp \left( - \mathrm{i} \alpha \cdot P - \overline{\theta} Q \right) |\varphi_I\rangle \eqend{,}
	\end{equation}
where $\theta = (\theta_1, \theta_2, \theta_3, \theta_4)$ are Grassmann numbers and
$\alpha^\mu = (\alpha^0, \vec \alpha)$ are commuting numbers.
As $P^\mu$ and $Q_\alpha$
commute, we can write this integral as
	\begin{eqnarray}
		|\varphi_I^{(BF)}\rangle & = & -\frac{1}{2V}\int \mathrm{d}^4 \theta \ \exp( -\overline{\theta} Q) \int \mathrm{d}^4 \alpha\,
\exp(- \mathrm{i} \alpha \cdot P) \ket{\varphi_I}\nonumber \\
&  =&  -Q_1 Q_2 Q_3 Q_4 |\varphi_I^{(B)}\rangle\eqend{,}
	\end{eqnarray}
where $|\varphi_I^{(B)}\rangle$ satisfies only the bosonic QLSCs:
	\begin{equation}
		|\varphi_I^{(B)}\rangle  = \frac{1}{2V}\int \mathrm{d}^4 \alpha\,\exp(- \mathrm{i} \alpha \cdot P) \ket{\varphi_I}\eqend{.}
	\end{equation}
We have shown that all states satisfying both the bosonic and fermionic QLSCs can be obtained in this manner.
The inner product $\langle \cdot|\cdot\rangle_{\mathcal{H}_{BF}}$ we have defined is
\begin{eqnarray}
\langle \varphi_1^{(BF)}|\varphi_2^{(BF)}\rangle_{\mathcal{H}_{BF}}
& = & - \frac{1}{2V}\int \mathrm{d}^4 \alpha \mathrm{d}^4 \theta\,\bra{\varphi_1}
\exp \left( -\mathrm{i} \alpha \cdot P - \overline{\theta} Q \right) \ket{\varphi_2}_{\mathcal{H}_0}\nonumber \\
& = & - \int \mathrm{d}^4 \theta \langle\varphi_1^{(B)}|\exp\left(
- \overline{\theta}Q\right)|\varphi_2^{(B)}\rangle_{\mathcal{H}_B}\eqend{.}
\end{eqnarray}		
In \ref{Appendix_C} we present an example of a state with two particles with non-zero momenta which satisfies all QLSCs.

\section{Summary and discussion} \label{summary}

In this paper we pointed out that there are fermionic linearization stability conditions  as well as bosonic ones in
$4$-dimensional $\mathcal{N}=1$ simple supergravity in the background of static $3$-torus space.  Then we showed that states satisfying
both fermionic and bosonic quantum linearization stability conditions (QLSCs) can be constructed by group-averaging over the supergroup of
global supersymmetry and spacetime translation symmetry.

States satisfying the bosonic QLSCs have infinite norm in the original Hilbert space.  This
infinity results from the infinite volume of the symmetry group generated by the LSCs.
Roughly speaking, this infinite volume is factored out in the group-averaging inner product.  It is interesting that
the inner product of states satisfying all QLSCs have zero norm in the Hilbert space of states
satisfying only the bosonic ones. The finite group-averaging inner product is obtained by factoring out zero in
this case.

As the bosonic QLSCs can be interpreted as a remnant of diffeomorphism invariance of the full generally covariant
theory, one should be able to interpret the fermionic QLSCs as a remnant of full local supersymmetry in the context
of canonical quantization~\cite{DEath1984,carroll1994}.
The bosonic QLSCs,
$H|\textrm{phys}\rangle = 0$ and $\vec{P}|\textrm{phys}\rangle =0$, have a natural physical picture.
These conditions imply that all physical states are invariant under spacetime translations.   This means that there is
no meaning in the position and time coordinates of an event relative to the background spacetime of static $3$-torus.  However, the physical
states still encode relative positions and relative time differences between two or more events.  Thus, the bosonic
QLSCs can be seen as a manifestation of Mach's principle in quantum general
relativity.  (See e.g.~\cite{barbour1982mach} for a discussion of Mach's principle in general relativity.)
On the other hand, it is not clear if there is a simple interpretation of the fermionic QLSCs.
It would be interesting to find one.

It would also be interesting to investigate whether there are analogues of LSCs in
String Theory.  A preliminary investigation in this direction~\cite{higuchi2001possible} did not find such analogues
in Bosonic String Theory, but since String Theory contains
General Relativity, we believe there should be analogues of
LSCs  in (Super)String Theory on any background spacetime with compact Cauchy surfaces.

The supergroup relevant to this work was a simple one with an abelian bosonic subgroup.
It would be interesting to investigate the group-averaging procedure for general supergroups and establish general
properties in analogy with the bosonic case studied in~\cite{giulini-marolf1999}.

\section*{Acknowledgements}
The work of L.\ S.\ was supported in part by a studentship from the Engineering and Physical Sciences Research
Council (EPSRC).

\appendix

\section{Modified transformation of the vierbein in general spacetime}\label{App-A}

In this appendix we show that one can modify the infinitesimal diffeomorphism transformation on $e_\mu^a$ by an infinitesimal local Lorentz transformation
so that the transform is proportional to $\nabla_\mu\zeta_\nu + \nabla_\nu \zeta_\mu$.
If we transform $e_\mu^a$ by diffeomorphism in the direction of $\zeta^\mu$ and by a local Lorentz transformation, we have
\begin{equation}
\delta^{(m)}e_\mu^a = \zeta^\rho\nabla_\rho e_\mu^a +e_\rho^a \nabla_\mu \zeta^\rho - s^{ab}e_{b\mu}\eqend{,}
\end{equation}
where $s^{ab}$ is anti-symmetric and where $\nabla_\rho e_\mu^a = \partial_\rho e_\mu^a - \Gamma_{\rho \mu}^\sigma e_\sigma^a$ with
the Levi-Civita connection $\Gamma_{\rho\mu}^\sigma$.
Our task is to choose $s^{ab}$ so that $\delta^{(m)}e_\mu^a$ is proportional to $\nabla_\mu \zeta_\nu + \nabla_\nu \zeta_\mu$.
To this end, we require
\begin{equation}
e_{a[\mu}\delta^{(m)}e_{\nu]}^a
= (\zeta^\rho \nabla_{\rho} e_{c[\nu}) e^c_{\mu]} + e_{c\rho} e_{[\mu}^c \nabla_{\nu]}\zeta^\rho -  s_{ab}e_{[\mu}^a e_{\nu]}^b  = 0\eqend{.}
\end{equation}
Since
\begin{equation}
D_\rho e_\nu^c = \nabla_\rho e_\nu^c + {\omega_\mu}^{cb}e_{b\nu}  = 0\eqend{,}
\end{equation}
we find
\begin{equation}
s_{ab}e^a_{[\mu} e^b_{\nu]} = -\zeta^\rho \omega_{\rho cb} e_{[\mu}^c e^b_{\nu]} + \nabla_{[\nu} \zeta_{\mu]}\eqend{.}
\end{equation}
Thus,
\begin{equation}
s^{ab} = - \zeta^\rho {\omega_\rho}^{ab} + \nabla_{[\nu} \zeta_{\mu]}e^{a\mu}e^{b\nu}\eqend{.}
\end{equation}
Then
\begin{eqnarray}
\delta^{(m)}e_\mu^a &  = &  \zeta^\rho \nabla_\rho e_\mu^a + e_\rho^a \nabla_\mu \zeta^\rho
+ \zeta^\rho{\omega_\rho}^{ab} + \frac{1}{2}(\nabla_\lambda\zeta_\nu - \nabla_{\nu}\zeta_\lambda)e^{a\lambda}e^{b\nu}e_{b\mu} \nonumber \\
& = & \zeta^\rho D_\rho e^a_\mu + e_\lambda^a \nabla_\mu \zeta^\lambda + \frac{1}{2}(\nabla_\lambda
\zeta_\mu - \nabla_\mu \zeta_\lambda)e^{a\lambda}
\nonumber \\
& = & \frac{1}{2}(\nabla_\lambda \zeta_\mu + \nabla_\mu \zeta_\lambda)e^{a\lambda}\eqend{.}
\end{eqnarray}

\section{Linearization stability condition for electrodynamics}\label{App-B1}

In this appendix we discuss the linearization stability condition for quantum electrodynamics in a static $3$-torus space. The Lagrangian density is
\begin{equation}
\mathcal{L} = - \frac{1}{4}F_{\mu\nu}F^{\mu\nu} + \overline{\psi}\gamma^\mu (\partial_\mu - \mathrm{i}e A_\mu)\psi - m\overline{\psi}\psi\eqend{,}
\end{equation}
where $F_{\mu\nu} =\partial_\mu A_\nu - \partial_\nu A_\mu$ and $\overline{\psi} = \mathrm{i}\psi^\dagger\gamma^0$.
The field equations are
\begin{eqnarray}
\frac{\delta S}{\delta A_\mu} &  =
& \partial_\nu F^{\nu\mu} - \mathrm{i}e \overline{\psi}\gamma^\mu\psi  = 0\eqend{,} \label{fieldeq-1}\\
\frac{\delta S}{\delta \psi^\dagger} & = &
 \mathrm{i}\gamma^0\left[\gamma^\mu (\partial_\mu - \mathrm{i}eA_\mu)\psi - m\psi\right]= 0\eqend{,} \label{fieldeq-2}\\
\frac{\delta S}{\delta\psi} & = &
(\partial_\mu + \mathrm{i}eA_\mu)\overline{\psi}\gamma^\mu  + m\overline{\psi} = 0\eqend{,} \label{fieldeq-3}
\end{eqnarray}
where $S$ is the action obtained by integrating $\mathcal{L}$ over spacetime.
We have defined the left-hand side of \eqref{fieldeq-2} and \eqref{fieldeq-3} as the left functional
derivative of the action $S$.

The field equations \eqref{fieldeq-2} and \eqref{fieldeq-3} imply that the current
$J_{(E)}^\mu = \mathrm{i}e\overline{\psi}\gamma^\mu \psi$ is conserved.  As a result, the total charge,
which is the
integral of $J_{(E)}^0$ over the space,
\begin{equation}
Q_{(E)}
= e \int\mathrm{d}^3\vec{x}\,\psi^\dagger\psi\eqend{,}
\end{equation}
is time independent.  This charge must vanish because the zeroth component of \eqref{fieldeq-1} implies
$\partial_i F^{i 0} = e \psi^\dagger \psi$ and hence
\begin{equation}
Q_{(E)} = \int \mathrm{d}^3\vec{x}\, \partial_i F^{i0}\eqend{,} \label{full-condition}
\end{equation}
which must vanish by Gauss's divergence theorem. (Recall our space is compact.)  At linearized level, the fields $A_\mu$
and $\psi$ are decoupled, and the condition $Q_{(E)}=0$ does not automatically arise in the linearized theory.  Hence
this condition has to be imposed as a linearization stability condition on the linearized solution for it
to extend to an exact solution.
The aim of this appendix is to illustrate how the argument in section~\ref{constraints} works in this simple model.

The action is invariant under the gauge transformation
$\delta A_\mu = \partial_\mu\Lambda$, $\delta\psi = \mathrm{i}e\Lambda\psi$ for any function $\Lambda$ of
the spacetime point.
The argument in section~\ref{constraints} leads to the conclusion that
\begin{eqnarray}
&&  \partial_\mu\Lambda(\partial_\nu F^{\nu\mu} - \mathrm{i}e\overline{\psi}\gamma^\mu\psi)
- \mathrm{i}e\Lambda \overline{\psi}\left[\gamma^\mu(\partial_\mu - \mathrm{i}eA_\mu)\psi - m\psi\right] \nonumber \\
&& - \mathrm{i}e \Lambda \left\{\left[ (\partial_\mu + \mathrm{i}eA_\mu)\overline{\psi}\right]\gamma^\mu + m\overline{\psi}\right\}\psi\nonumber \\
&&  = \partial_\mu J_{(\Lambda)}^\mu\left[A,\psi\right] \eqend{,}  \label{QED-identity}
\end{eqnarray}
for some current $J_{(\Lambda)}^\mu\left[A,\psi\right]$ for any $\Lambda$.  We indeed find that this
equation is satisfied with
\begin{equation}
J_{(\Lambda)}^\mu\left[A,\psi\right] = \Lambda(\partial_\nu F^{\nu \mu} - \mathrm{i}e\overline{\psi}\gamma^\mu \psi)\eqend{.}
\end{equation}

The identity \eqref{QED-identity} is satisfied order by order in the fields $A_\mu$ and $\psi$, i.e.
\begin{eqnarray}
& \partial_\mu \Lambda \partial_\nu F^{\nu\mu}  = \partial_\mu J_{(\Lambda)}^{(1)\mu}\left[A\right]\eqend{,}
\label{first-order-A}\\
& -\mathrm{i}e \partial_\mu\Lambda \overline{\psi}\gamma^\mu \psi  -
\mathrm{i}e\Lambda\overline{\psi}\gamma^\mu\partial_\mu\psi
-  \mathrm{i}e\Lambda (\partial_\mu\overline{\psi})\gamma^\mu\psi
= \partial_\mu J_{(\Lambda)}^{(2)\mu}\left[A,\psi\right]\eqend{,} \label{second-order-A}
\end{eqnarray}
where
\begin{eqnarray}
J_{(\Lambda)}^{(1)\mu}[A] & = \Lambda\partial_\nu F^{\nu\mu}\eqend{,}\\
J_{(\Lambda)}^{(2)\mu}[A,\psi] & = - \mathrm{i}e\Lambda \overline{\psi}\gamma^\mu \psi\eqend{.}
\end{eqnarray}
Now, if $\Lambda(x)=\lambda$ is constant, then the left-hand side of \eqref{first-order-A} vanishes.  Hence
the current $J^{(1)\mu}_{(\lambda)}[A] = \lambda \partial_\nu F^{\nu \mu}$ is conserved for any field configuration
$A_\mu$.  In particular, we can smoothly deform $A_\mu$ to zero in the far future or past of any given Cauchy surface while keeping the value of the conserved charge
	\begin{equation}
		Q^{(1)}_{(E)}[A] = \int \mathrm{d}^3\vec{x}\,\, \partial_i F^{i 0} \eqend{,}
	\end{equation}
unchanged. It follows that $Q^{(1)}_{(E)}[A] = 0$ for any field $A_\mu$.  Of course, one can readily verify this fact by Gauss's divergence theorem, as we observed before.

Now, consider solving the field equations order by order by letting
		\begin{eqnarray}
		A_\mu & = & A^{(1)}_\mu + A^{(2)}_{\mu} + \cdots\eqend{,}\\
		\psi & = & \psi^{(1)} + \psi^{(2)}+\cdots\eqend{.}
\end{eqnarray}
	Then, with the definition $F^{(I)}_{\mu\nu}:=\partial_\mu A_\nu^{(I)} - \partial_\nu A_\mu^{(I)}$, $I=1,2$, the field equations~\eqref{fieldeq-1}-\eqref{fieldeq-3} become
\begin{eqnarray}
 \partial_\nu F^{(1)\nu\mu} = 0\eqend{,}\\
\mathrm{i}\gamma^0\left[\gamma^\mu\partial_\mu \psi^{(1)} - m\psi^{(1)}\right] = 0\eqend{,}\\
 \partial_\mu \overline{\psi^{(1)}}\gamma^\mu + m\overline{\psi^{(1)}} = 0\eqend{,}\\
\partial_\nu F^{(2)\nu\mu}-
\mathrm{i}e \overline{\psi^{(1)}}\gamma^\mu \psi^{(1)} = 0\eqend{.} \label{2nd-field-eq}
\end{eqnarray}
Then, equations \eqref{first-order-A} and \eqref{second-order-A} imply that the current
\begin{equation}
J_{(\Lambda)}^{(1)\mu}[A^{(2)}] + J_{(\Lambda)}^{(2)\mu}[A^{(1)},\psi^{(1)}]
= \Lambda (\partial_\nu F^{(2)\nu\mu} - \mathrm{i}e\overline{\psi^{(1)}}\gamma^\mu\psi^{(1)})
\eqend{,}
\end{equation}
is conserved for any $\Lambda(x)$.  This implies that the corresponding charge must vanish, which
is rather obvious in this example because, in fact,
$J^{(1)\mu}_{(\Lambda)}[A^{(2)}] + J^{(2)\mu}_{(\Lambda)}[A^{(1)},\psi^{(1)}] = 0$ by \eqref{2nd-field-eq}.
In particular, if we define
\begin{equation}
Q_{(E)}^{(2)}[\psi^{(1)}] =  - e\int\mathrm{d}^3\vec{x}\, \psi^{(1)\dagger}\psi^{(1)}\eqend{,}
\end{equation}
then, since $e\lambda \psi^{(1)\dagger}\psi^{(1)}
= \mathrm{i}e\lambda \overline{\psi^{(1)}}\gamma^0\psi^{(1)} = - J^{(2)0}_{(\lambda)}[A^{(1)},\psi^{(1)}]$,
we find
\begin{equation}
Q_{(E)}^{(2)}[\psi^{(1)}] = - Q^{(1)}_{(E)}[A^{(2)}] = 0\eqend{,}
\end{equation}
which is the linearization stability condition for electrodynamics in static torus space.  In this example, it is the
linearization of the condition \eqref{full-condition} in the exact theory.

\section{Proof of identities \eqref{interesting-identity} and \eqref{standard-identity}} \label{App-B}

We first note that the matrices $T^B_{ij}$, $B=1,2$, are diagonal with $T\indices{^B^i_k} = \delta^i_k q^B_k$, where
$q^1_1 = - q^1_2 = 1/\sqrt{2}$, $q^1_3 =0$. $q^2_1=q^2_2 = 1/\sqrt{6}$ and $q^2_3 = -2/\sqrt{6}$.   Hence
\begin{eqnarray}
X & := & \sum_{B=1}^2\sum_{A=1}^5 \sum_{A'=1}^5
T^A_{ij} T\indices{^B^i_k}T^{A'}_{i'j'}T\indices{^B^{i'}_{k'}}\gamma^j\gamma^k \gamma^{k'}\gamma^{j'}c_{PA}c_{PA'} \nonumber \\
& = & \sum_{k=1}^3\sum_{k'=1}^3
\sum_{A=1}^5\sum_{A'=1}^5 T_{kj}^A T^{A'}_{k'j'}\sum_{B=1}^2 q^B_k q^B_{k'}\gamma^j\gamma^k\gamma^{k'}\gamma^{j'}c_{PA}c_{PA'}
\eqend{.} \label{B1}
\end{eqnarray}
By a component-by-component calculation we find
\begin{equation}
\sum_{B=1}^2 q^B_k q^B_{k'} = \delta_{kk'} - \frac{1}{3}\eqend{.}
\end{equation}
By substituting this formula into \eqref{B1} we obtain
\begin{eqnarray}
X & = &
\sum_{A=1}^5\sum_{A'=1}^5 T\indices{^A^k_j}T^{A'}_{kj'}\gamma^j\gamma^{j'}c_{PA}c_{PA'}
-  \frac{1}{3}
\sum_{A=1}^5\sum_{A'=1}^5 T_{kj}^A T^{A'}_{kj'}\gamma^j\gamma^k \gamma^{k'}\gamma^{j'}c_{PA}c_{PA'\eqend{.}}\nonumber \\
\end{eqnarray}
The second sum vanishes because the matrices $T^A_{ij}$ are symmetric and traceless so that
$T^A_{kj}\gamma^j \gamma^k = T^A_{jk}(\delta^{jk}+\gamma^{jk}) = 0$.  Since the first sum is of the form $S_{jj'}\gamma^j\gamma^{j'}$ where
$S_{jj'}$ is symmetric, we may replace $\gamma^{j}\gamma^{j'}$ by $\delta^{jj'}$.  Hence
\begin{equation}
X  = \sum_{A=1}^5 \sum_{A'=1}^5 T\indices{^A^k_j}T^A_{kj'}\delta^{jj'}c_{PA}c_{PA'}
=  \sum_{A=1}^5 \delta^{AA'}c_{PA}c_{PA'}
= \sum_{A=1}^5 (c_{PA})^2\eqend{,}
\end{equation}
which is equation \eqref{interesting-identity}.

To show \eqref{standard-identity} we note that, since $u^\pm(\vec{k})$ is a simultaneous eigenspinor of
$\gamma^0\hat{k}\cdot \vec{\gamma}$ and $\gamma_5$ with eigenvalues $-1$ and $\pm 1$, respectively, with the normalization
$u^{\pm\dagger}(\vec{k})u^{\pm}(\vec{k}) = 2k$,
the matrix $M_{\alpha\beta} = u^{\pm}(\vec{k})_\alpha u^{\pm *}(\vec{k})_\beta$ is the product of the projection operator onto
the eigenspace of $\gamma^0 \hat{k}\cdot\vec{\gamma}$ with eigenvalue $-1$ and that of $\gamma_5$ with eigenvalue $\pm 1$ (multiplied by $2k$) since
these projection operators commute.  Hence,
we have
\begin{eqnarray}
u^{\pm}(\vec{k})_\alpha u^{\pm*}(\vec{k})_\beta & = & \frac{k}{2}\left[1 - \gamma^0\hat{k}\cdot\vec{\gamma})(1\pm \gamma_5)\right]_{\alpha\beta}
\nonumber \\
& = & \frac{1}{2}\left[(1\pm \gamma_5) k\cdot\gamma\gamma^0\right]_{\alpha\beta}\eqend{,}
\end{eqnarray}
where we have used $k\cdot \gamma = k(- \gamma^0 + \hat{k}\cdot \vec{\gamma})$ so that
$k(1-\gamma^0 \hat{k}\cdot \vec{\gamma}) = k\cdot \gamma \gamma^0$.
Hence
\begin{eqnarray}
\left[ \epsilon^{\pm*}(\vec{k})\cdot\vec{\gamma} u^\pm(\vec{k})\right]_\alpha
\left[\epsilon^{\pm}(\vec{k})\cdot\vec{\gamma} u^{\pm*}(\vec{k})\right]_\beta \nonumber\\
= \frac{1}{2}\left[ (1\mp\gamma_5) k\cdot\gamma (\epsilon^{\pm *}(\vec{k})\cdot\vec{\gamma})(\epsilon^{\pm}(\vec{k})\cdot\vec{\gamma})
\gamma^0\right]_{\alpha\beta}
\eqend{,} \label{intermediate-standard}
\end{eqnarray}
where we have used the fact that $k\cdot \gamma$, $\gamma^0$ and $\gamma_5$ all anti-commute with
$\epsilon^{\pm}(\vec{k})\cdot \vec{\gamma}$ and $\epsilon^{\pm *}(\vec{k})\cdot \vec{\gamma}$.
Now the identities
$(\epsilon^{\pm *}(\vec{k})\cdot\vec{\gamma})(\epsilon^{\pm}(\vec{k})\cdot\vec{\gamma}) = 1\pm \gamma^0 \hat{k}\cdot\vec{\gamma}\gamma_5$ and
$k\cdot\gamma\gamma^0 \hat{k}\cdot\vec{\gamma} =  k\cdot \gamma$
imply $ k\cdot\gamma (\epsilon^{\pm *}(\vec{k})\cdot\gamma)(\epsilon^{\pm}(\vec{k})\cdot\gamma) =  k\cdot\gamma (1\pm \gamma_5)$.
By substituting this identity into \eqref{intermediate-standard} we find
\begin{equation}\left[ \epsilon^{\pm*}(\vec{k})\cdot\vec{\gamma} u^\pm(\vec{k})\right]_\alpha
\left[\epsilon^{\pm}(\vec{k})\cdot\vec{\gamma} u^{\pm*}(\vec{k})\right]_\beta
= \left[ (1 \mp \gamma_5)k\cdot\gamma \gamma^0\right]_{\alpha\beta}\eqend{.}
\end{equation}
Finally, by noting that $[(1\mp \gamma_5)k\cdot\gamma \gamma^0]_{\beta\alpha} = [(1\pm \gamma_5)k\cdot\gamma \gamma^0]_{\alpha\beta}$ we
find \eqref{standard-identity}.

\section{Zero-momentum sector of the gravitino field} \label{fermion-zero-mom-sector}

In this appendix we provide some details of the (indefinite-metric) Hilbert space describing the zero-momentum sector of
the gravitino field.  Some of the material presented here can be found in \cite{henneaux1994quantization}.  We make the discussion general and treat the fermionic algebra where the self-adjoint operators
$\eta_\alpha$, $\alpha=1,2,\ldots,2M$ and $\eta^{(+)}_\alpha$, $\alpha=1,2,\ldots,2N$ satisfying
\begin{equation}\label{eq:fermionic-algebra}
[ \eta_\alpha,\eta_\beta]_+ = - \delta_{\alpha\beta}\eqend{,}\ \ 
[ \eta^{(+)}_\alpha,\eta^{(+)}_\beta]_{+} = \delta_{\alpha\beta}\eqend{,}\ \ \ 
[\eta_\alpha,\eta^{(+)}_\beta]_{+} = 0\eqend{.}
\end{equation}
The zero-momentum sector of the gravitino field corresponds to $M=2$ and $N=4$.

We represent this algebra as follows.  We define annihilation-type operators as
\begin{eqnarray}
d_a & : = & \frac{1}{\sqrt{2}}(\eta_{2a-1} + \mathrm{i}\eta_{2a})\eqend{,}\ 
a=1,2,\ldots, M\eqend{,}\\
d^{(+)}_a & : = & \frac{1}{\sqrt{2}}(\eta^{(+)}_{2a-1} + \mathrm{i}\eta^{(+)}_{2a})\eqend{,}\ 
a=1,2,\ldots, N\eqend{.}
\end{eqnarray}
These operators and their adjoint, the creation-type operators, have the following non-zero anti-commutators:
\begin{equation}
[ d_a,d_b^\dagger]_+ = -\delta_{ab}\eqend{,}\ [d^{(+)}_a,d^{(+)\dagger}_b]_{+}  = \delta_{ab}\eqend{.}
\end{equation}
We define the state $|0_{0F}\rangle$ by requiring
$d_a|0_{0F}\rangle = d^{(+)}_a|0_{0F}\rangle = 0$ for all $a$ and  $\langle 0_{0F}|0_{0F}\rangle = 1$.
Then the $2^{M+N}$-dimensional (indefinite-metric) Hilbert space representing the algebra \eqref{eq:fermionic-algebra}
is spanned by the following orthogonal states:
\begin{equation}\label{eq:general-state}
|\left\{n\right\},\left\{ n^{(+)}\right\}\rangle = \prod_{a=1}^M (d_a^\dagger)^{n_a}
\prod_{b =1}^N (d_b^{(+)\dagger})^{n_b}|0_{0F}\rangle\eqend{,}
\end{equation}
where $n_a$ and $n^{(+)}_b$ are either $0$ or $1$ and where
$\left\{ n\right\} = \left\{ n_1,n_2,\ldots,n_M\right\}$ and
$\left\{ n^{(+)}\right\} = \left\{ n_1^{(+)},n_2^{(+)},\ldots,n_N^{(+)}\right\}$.  One readily finds
\begin{equation}\label{eq:general-norm}
\langle \left\{ n\right\},\left\{ n^{(+)}\right\}|\left\{ n\right\},\left\{ n^{(+)}\right\}\rangle
=\left\{ \begin{array}{cc} 1 & \textrm{if}\ n_1+n_2+\cdots+n_M\ \textrm{is even},\\
-1  &  \textrm{if}\ n_1+n_2+\cdots + n_N\ \textrm{is odd}. \end{array} \right.
\end{equation}

If $\Omega$ is an anti-self-adjoint operator satisfying $\Omega^\dagger = - \Omega$, 
then the operator $\exp\Omega$, which can be defined as a power series,
is unitary in the sense that it preserves the inner product.  
Of particular interest are the following unitary operators:
\begin{eqnarray}
U_{\alpha\beta}(\theta) & := & \exp\left(  \frac{\theta}{2}\eta_\alpha \eta_\beta\right)
= \cos\frac{\theta}{2} + \eta_\alpha \eta_\beta \sin\frac{\theta}{2}
\eqend{,}\ \ 
\alpha\neq \beta\eqend{,}\\
V_{\alpha\beta}(\theta) & := & \exp\left(  \frac{\theta}{2}\eta_\alpha \eta^{(+)}_\beta\right)
= \cosh \frac{\theta}{2} + \eta_\alpha \eta^{(+)}_\beta \sinh \frac{\theta}{2}
\eqend{,}\\
W_{\alpha\beta}(\theta) & := & \exp\left(  \frac{\theta}{2}\eta^{(+)}_\alpha\eta^{(+)}_\beta\right)
= \cos\frac{\theta}{2} + \eta_\alpha^{(+)}\eta_\beta^{(+)} \sin\frac{\theta}{2}
\eqend{,}\ \ 
\alpha\neq \beta\eqend{.}
\end{eqnarray}
These unitary operators act on $\eta_\alpha$ and $\eta^{(+)}_\alpha$ as follows:
\begin{eqnarray}
U_{\alpha\beta}(\theta) \eta_\alpha U_{\alpha\beta}(\theta)^{\dagger}
& = & \eta_\alpha\cos\theta + \eta_\beta \sin\theta\eqend{,}\\
W_{\alpha\beta}(\theta)\eta^{(+)}_\alpha W_{\alpha\beta}(\theta)^{\dagger}
& = & \eta^{(+)}_\alpha\cos\theta - \eta^{(+)}_\beta \sin\theta\eqend{,}\\
V_{\alpha\beta}(\theta)\eta_\alpha V_{\alpha\beta}(\theta)^{\dagger}
& = & \eta_\alpha\cosh \theta + \eta^{(+)}_\beta \sinh\theta\eqend{,}\\
V_{\alpha\beta}(\theta)\eta^{(+)}_\beta V_{\alpha\beta}(\theta)^{\dagger}
& = & \eta^{(+)}_\beta\cosh \theta + \eta_\alpha \sinh\theta\eqend{.}
\end{eqnarray}
Let us define $Y:= \prod_{\alpha=1}^{2M} \eta_\alpha \prod_{\beta=1}^{2N}\eta^{(+)}_\beta$.  Then,  $Y$ is unitary,
i.e.\ $Y^\dagger = Y^{-1}$.  The operator $W_\alpha:=Y\eta^{(+)}_\alpha$, $\alpha=1,2,\ldots,2N$, is also unitary.
We find $W_\alpha \eta^{(+)}_\alpha W_\alpha^\dagger = - \eta^{(+)}_\alpha$ whereas
$W_\alpha \eta^{(+)}_\beta W_\alpha^\dagger = \eta^{(+)}_\beta$ if $\beta\neq\alpha$ and
$W_\alpha\eta_\beta W_\alpha^\dagger = \eta_\beta$ for all $\beta$.  

Since any product of
$U_{\alpha\beta}(\theta)$, $W_{\alpha\beta}(\theta)$, $V_{\alpha}(\theta)$ and $W_\alpha$ is unitary, we can conclude the
following.  Suppose that $M \leq N$ and let
\begin{equation}
\tilde{X}_\alpha = {A_\alpha}^\beta \eta_\beta \cosh\theta_\alpha - {C_\alpha}^\beta \eta^{(+)}_\beta \sinh\theta_\alpha
\eqend{,}
\end{equation}
where the index $\alpha=1,2,\ldots, 2M$, is not summed over, and where $({A_\alpha}^\beta)$ and $({C_\alpha}^\beta)$
are a $2M\times 2M$ matrix and a $2M\times 2N$ matrix, respectively, satisfying
${A_{\alpha_1}}^{\beta_1}{A_{\alpha_2}}^{\beta_2}\delta_{\beta_1\beta_2} = 
{C_{\alpha_1}}^{\beta_1}{C_{\alpha_2}}^{\beta_2}\delta_{\beta_1\beta_2} = \delta_{\alpha_1\alpha_2}$ and
$\textrm{det}({A^\alpha}_\beta) = 1$.\footnote{The unitarity of the operators $W_\alpha$ allows us to have
$\textrm{det}({C^\alpha}_\beta) = -1$.}
Then there is a unitary operator $U$ such that
$\tilde{X}_\alpha = U\eta_\alpha U^\dagger$.

\section{Example of a state satisfying all quantum linearization stability conditions} \label{Appendix_C}

In this appendix we present a concrete example of a state satisfying all fermionic as well as bosonic QLSCs. We consider an example where there are two particles, one going
in the positive $x$-direction and the other in the negative $x$-direction with momentum $k > 0$ and both with positive helicity. In this sector of the theory $M^2 = 4k$.

To find the non-zero-momentum contribution to the supercharge in this sector we need to find the spinors $u^{+}(\vec{k})$ with $\vec{k}= \pm k\hat{e}_1$, where $\hat{e}_1$ is the unit vector
in the $x$-direction.  It is
an eigenspinor of $\gamma^0\gamma^1$ and $\mathrm{i}\gamma^2\gamma^3$ with eigenvalue $-1$ for both and, hence, an eigenspinor of
$\gamma_5 = \mathrm{i}\gamma^0\gamma^1\gamma^2\gamma^3$ with eigenvalue $1$.  It is normalized as
$u^{+\dagger}(\pm k\hat{e}_1)u^{+}(\pm k\hat{e}_1) = 2k$.  We find from \eqref{gamma-choice}
\begin{equation}
\gamma^0 \gamma^1  = \left( \begin{array}{cc} 0 & - 1\\ - 1 & 0\end{array}\right)\eqend{,}\ \ \
\mathrm{i}\gamma^2\gamma^3  =\left( \begin{array}{cc} \sigma_2 & 0 \\ 0 & \sigma_2\end{array}\right) \eqend{.}
\end{equation}
Hence we can take
\begin{equation}
u^+(k\hat{e}_1)  =  \sqrt{\frac{k}{2}}\left( \begin{array}{c} 1 \\ -\mathrm{i} \\ 1 \\ -\mathrm{i} \end{array}\right)\eqend{,}\ \ \
u^{+*}(k\hat{e}_1)  = \sqrt{\frac{k}{2}} \left( \begin{array}{c} 1 \\ \mathrm{i} \\ 1 \\ \mathrm{i}\end{array}\right)\eqend{.}
\end{equation}
Now, $\{ \hat{e}_2, \hat{e}_3, \hat{e}_1 \}$ form a right handed basis so that we can choose
\begin{equation}
\epsilon^{+}(k\hat{e}_1) \cdot \vec{\gamma} =\frac{1}{\sqrt{2}}( \gamma^2 + \mathrm{i}\gamma^3) =
\frac{1}{\sqrt{2}}\gamma^2(1 + \mathrm{i}\gamma^2\gamma^3)\eqend{.}
\end{equation}
Since $u^{+*}(k\hat{e}_1)$ is an eigenspinor of $\mathrm{i}\gamma^2\gamma^3$ with eigenvalue $+1$, we have
\begin{equation}
\epsilon^{+}(k\hat{e}_1)\cdot\vec{\gamma} u^{+*}(k\hat{e}_1) = \sqrt{k}\gamma^2
\left( \begin{array}{c} 1 \\ \mathrm{i} \\ 1 \\ \mathrm{i}\end{array}\right)
= \sqrt{k}\left( \begin{array}{c} \mathrm{i} \\ 1 \\ \mathrm{i} \\ 1\end{array}\right)\eqend{,}
\end{equation}
and
\begin{equation}
\epsilon^{+*}(k\hat{e}_1)\cdot\vec{\gamma} u^{+}(k\hat{e}_1) = \sqrt{k} \gamma^2
\left( \begin{array}{c} 1 \\ - \mathrm{i} \\ 1 \\ - \mathrm{i}\end{array}\right)
= \sqrt{k}\left( \begin{array}{c} -\mathrm{i} \\ 1 \\ - \mathrm{i} \\ 1\end{array}\right)
\eqend{.}
\end{equation}

The spinor $u^{+}(-k\hat{e}_1)$ is an eigenspinor of $\gamma^0\gamma^1$ and $\mathrm{i}\gamma^2\gamma^3$ with eigenvalue $+1$ for both.
We can choose
\begin{equation}
\epsilon^{+}(-k\hat{e}_1)\cdot \vec{\gamma} = \frac{1}{\sqrt{2}}\gamma^2 (1 - \mathrm{i}\gamma^2\gamma^3)\eqend{.}
\end{equation}
Proceeding similarly as above, we find
\begin{equation}
\epsilon^{+}(-k\hat{e}_1)\cdot\vec{\gamma} u^{+*}(-k\hat{e}_1) = \sqrt{k}\gamma^2
\left( \begin{array}{c} 1 \\ \mathrm{i} \\ -1 \\ -\mathrm{i}\end{array}\right)
= \sqrt{k} \left( \begin{array}{c} -\mathrm{i} \\ -1 \\ \mathrm{i} \\ 1\end{array}\right)\eqend{,}
\end{equation}
and
\begin{equation}
\epsilon^{+*}(-k\hat{e}_1)\cdot\vec{\gamma} u^{+}(-k\hat{e}_1) = \sqrt{k}\gamma^2
\left( \begin{array}{c} 1 \\ -\mathrm{i} \\ -1 \\ \mathrm{i}\end{array}\right)
= \sqrt{k} \left( \begin{array}{c} \mathrm{i} \\ -1 \\ -\mathrm{i} \\ 1\end{array}\right)\eqend{,}
\end{equation}
Thus, the relevant part of the $\vec{k}\neq 0$ contribution to the supercharge
\begin{equation} \fl
\qquad \widehat{Q} =  -  \frac{\mathrm{i}}{2}
\sum_{\vec k \neq 0} \sum_{\lambda = \pm} \left[ \epsilon^\lambda(\vec k)\cdot\gamma u^{\lambda*}(\vec k) a_\lambda (\vec{k})
b_{\lambda}^\dagger(\vec k) - \epsilon^{\lambda *}(\vec k)\cdot\gamma u^\lambda(\vec k) a_\lambda^\dagger(\vec k) b_\lambda (\vec k) \right]
\eqend{,}
\end{equation}
is, with the notation $a_{(\pm)} = a_+(\pm k\hat{e}_1)$ and  $b_{(\pm)}=b_{+}(\pm k\hat{e}_1)$,
\begin{equation}
\widehat{Q} = \frac{\sqrt{k}}{2} \left( \begin{array}{c} \left[ a_{(+)}b_{(+)}^\dagger + a_{(+)}^\dagger b_{(+)}\right]
- \left[ a_{(-)}b_{(-)}^\dagger + a_{(-)}^\dagger b_{(-)}\right] \\
- \mathrm{i}\left[ a_{(+)}b_{(+)}^\dagger - a_{(+)}^\dagger b_{(+)}\right]
+ \mathrm{i}\left[ a_{(-)}b_{(-)}^\dagger - a_{(-)}^\dagger b_{(-)}\right] \\
\left[ a_{(+)}b_{(+)}^\dagger + a_{(+)}^\dagger b_{(+)}\right]
+ \left[ a_{(-)}b_{(-)}^\dagger + a_{(-)}^\dagger b_{(-)}\right] \\
- \mathrm{i}\left[ a_{(+)}b_{(+)}^\dagger - a_{(+)}^\dagger b_{(+)}\right]
- \mathrm{i}\left[ a_{(-)}b_{(-)}^\dagger - a_{(-)}^\dagger b_{(-)}\right] \end{array}\right)\eqend{.}
\end{equation}

Proceeding as in section~\ref{fermionic-constraints} to construct a state satisfying the fermionic QLSCs we define
\begin{equation}
a_1 = \frac{\sqrt{2}}{M}( Q^{(0)}_1 + \mathrm{i} Q^{(0)}_2 )\eqend{,}\ \
a_2  = \frac{\sqrt{2}}{M}( Q^{(0)}_3 + \mathrm{i} Q^{(0)}_4 )\eqend{,}
\end{equation}
as before and also
\begin{eqnarray}
b_1 = \frac{1}{\sqrt{2 k} } (\widehat{Q}_1 + \mathrm{i} \widehat{Q}_2) = \frac{1}{\sqrt{2}} ( a_{(+)} b_{(+)}^\dagger - a_{(-)} b_{(-)}^\dagger ) \eqend{,} \\
b_2 = \frac{1}{\sqrt{2 k} } (\widehat{Q}_3 + \mathrm{i} \widehat{Q}_4) = \frac{1}{\sqrt{2}} ( a_{(+)} b_{(+)}^\dagger + a_{(-)} b_{(-)}^\dagger ) \eqend{,}
\end{eqnarray}
in the relevant subspace.
Then in the two-particle sector, $M^2 = 4k$, it is easy to see that
	\begin{equation}
		[a_i,a_j^\dagger]_{+} = - \delta_{ij} \eqend{,} \quad  [b_i,b_j^\dagger]_{+} = \delta_{ij} \eqend{.}
	\end{equation}
	
We define a state $\ket{\varphi^{(B)}}$ obeying the bosonic QLSCs as follows. We take the bosonic zero-momentum sector wave function to be given by the Gaussian
	\begin{equation}
		\Psi(c, \vec c) = N \mathrm{e}^{- c^2} \mathrm{e}^{- \vec c ^2} \eqend{,}
	\end{equation}
where $N$ is a constant,
and the non-zero-momentum part of the state contains two positive helicity gravitons with momentum $\pm k \hat{e}_1$. Then the state obeying the bosonic QLSCs by group-averaging is given by \eqref{invariant-state}, which reads
	\begin{equation} \fl
		\qquad \ket{\varphi^{(B)}} = \int \frac{\mathrm{d}^5\vec{p}}{(2\pi)^5}
\left[ f^{(+)}(\vec{p})
\mathrm{e}^{-\mathrm{i}E(\vec{p}) c + \mathrm{i}\vec{p}\cdot \vec{c}}
+ f^{(-)}(\vec{p})
\mathrm{e}^{\mathrm{i}E(\vec{p})c + \mathrm{i}\vec{p}\cdot\vec{c}}\right]\otimes \left( a_{(+)}^\dagger a_{(-)}^\dagger \ket{0} \right) \eqend{,} \label{eq:mom-decomposition}
	\end{equation}
where the state $\ket{0}$ is the Fock vacuum in the non-zero-momentum sector, $E(\vec p) = \sqrt{ \vec p^2 + M^2}$ and
	\begin{equation}
		f^{(\pm)}(\vec p) = 
\frac{\pi^3 N\mathrm{e}^{-M^2/4}\mathrm{e}^{- \vec p ^2 / 2}}{2\sqrt{\vec{p}^2 + M^2}} \eqend{.}
	\end{equation}
The norm of this state in the Hilbert space $\mathcal{H}_B$ is then given by \eqref{GA-inner-product}, which reads
	\begin{equation}
		\braket{\varphi^{(B)}}{\varphi^{(B)}}_{\mathcal{H}_B} = 
\frac{\pi^3}{12}|N|^2\mathrm{e}^{-M^2/2}
\int^\infty_0 \mathrm{d} p \ \frac{p^4 e^{- p^2}}{\sqrt{p^2 + M^2}} < \infty\eqend{.}
	\end{equation}
We choose the normalisation constant $N$ so that $\braket{\varphi^{(B)}}{\varphi^{(B)}}_{\mathcal{H}_B} = 1$. 

Now we need to consider the fermionic zero-momentum sector.  
As we stated in section~\ref{fermionic-constraints}, we need to 
consider the tensor product of $\mathcal{H}_B$ and the $64$-dimensional Hilbert space $\mathcal{H}_{0F}$
describing the fermionic zero-momentum sector.  We call this tensor product $\mathcal{H}_B$ as in 
section~\ref{fermionic-constraints}.  As in \ref{fermion-zero-mom-sector} let $|0_{0F}\rangle$ be the normalized state 
annihilated by
$d_1 = (\eta_1+\mathrm{i}\eta_2)/\sqrt{2}$, $d_2 = (\eta_3+\mathrm{i}\eta_4)/\sqrt{2}$,
$d_1^B = (\eta_1^B+\mathrm{i}\eta_2^B)/\sqrt{2}$ and $d_2^B=(\eta_3^B + \mathrm{i}\eta_4^B)/\sqrt{2}$,
$B=1,2$.  Let $\bar{a}_1$ and $\bar{a}_2$ be the operators $a_1$ and $a_2$, respectively, restricted to the
eigenspace of $c_P$ and $c_{PA}$ with eigenvalues $p_0$ and $p_A$, respectively. 
Define $|p_0,\vec{p}\rangle \in \mathcal{H}_{0F}$ by
$|p_0,\vec{p}\rangle := C_{p_0,\vec{p}}\,\bar{a}_1\bar{a}_2 d_2^\dagger d_1^\dagger|0_{0F}\rangle$, where
$C_{p_0,\vec{p}}$ is a normalization constant such that $\langle p_0,\vec{p}|p_0,\vec{p}\rangle_{\mathcal{H}_{0F}}=1$.
Then, $\bar{a}_1|p_0,\vec{p}\rangle = \bar{a}_2|p_0,\vec{p}\rangle = 0$ because
$\bar{a}_1^2 =\bar{a}_2^2 = 0$ and $\bar{a}_1\bar{a}_2 = - \bar{a}_2\bar{a}_1$.  Now, we let
\begin{eqnarray}
\fl \ket{\varphi^{(B)}} =  \int \frac{\mathrm{d}^5\vec{p}}{(2\pi)^5}
\left[ f^{(+)}(\vec{p})
\mathrm{e}^{-\mathrm{i}E(\vec{p}) c + \mathrm{i}\vec{p}\cdot \vec{c}}\otimes |E(\vec{p}),\vec{p}\rangle
+ f^{(-)}(\vec{p})
\mathrm{e}^{\mathrm{i}E(\vec{p})c + \mathrm{i}\vec{p}\cdot\vec{c}}\otimes |-E(\vec{p}),\vec{p}\rangle \right]
\nonumber \\
\ \ \ \ \otimes \left( a_{(+)}^\dagger a_{(-)}^\dagger \ket{0} \right) \eqend{.}
	\end{eqnarray}
Then,
$a_1 \ket{\varphi^{(B)}} = a_2 \ket{\varphi^{(B)}} = 0$ and
$\langle \varphi^{(B)}|\varphi^{(B)}\rangle_{\mathcal{H}_B} = 1$.
Notice also that because $\ket{\varphi^{(B)}}$ contains only gravitons, it obeys 
$b^\dagger_1 \ket{\varphi^{(B)}} = b^\dagger_2 \ket{\varphi^{(B)}} = 0$. 

In the notation of 
section~\ref{fermionic-constraints} we can thus take $\ket{\varphi^{(B)}} \propto \ket{0011}$. Indeed, by the 
group-averaging procedure described in section~\ref{fermionic-constraints}, 
we obtain a state $\ket{\varphi^{(BF)}} \in \mathcal{H}_{BF}$ obeying the bosonic and fermionic QLSCs 
by applying $-Q_1 Q_2 Q_3 Q_4$. That is, 
	\begin{eqnarray}
		\ket{\varphi^{(BF)}} & = & - Q_1 Q_2 Q_3 Q_4 \ket{\varphi^{(B)}} \\
		& = & \frac{1}{4} (Q_1 - \mathrm{i} Q_2)(Q_1 + \mathrm{i} Q_2)(Q_3 - \mathrm{i} Q_4)(Q_3 + \mathrm{i} Q_4) \ket{\varphi^{(B)}} \\
		& = & \frac{M^4}{16} (a_1^\dagger + b_1^\dagger)(a_1 + b_1)(a_2^\dagger + b_2^\dagger)(a_2 + b_2)  \ket{\varphi^{(B)}} \eqend{.}
	\end{eqnarray}
By writing $\ket{\varphi^{(B)}} = a_{(+)}^\dagger a_{(-)}^\dagger \ket{\Psi^{(B)}}$ this can be evaluated as
	\begin{eqnarray}
		\ket{\varphi^{(BF)}} 
		&=& \frac{M^4}{16} \left[  a_{(+)}^\dagger a_{(-)}^\dagger|\Psi^{(B)} \rangle
- \frac{1}{\sqrt{2}}( a_1^\dagger - a_2^\dagger) a_{(+)}^\dagger b_{(-)}^\dagger|\Psi^{(B)} \rangle \right. \nonumber \\
&& \ \ \ \ \left.  +
\frac{1}{\sqrt{2}}( a_1^\dagger + a_2^\dagger) b_{(+)}^\dagger a_{(-)}^\dagger|\Psi^{(B)} \rangle
- a_{1}^\dagger a_{2}^\dagger b_{(+)}^\dagger b_{(-)}^\dagger |\Psi^{(B)} \rangle \right]\eqend{.}
	\end{eqnarray}
In particular, by the group-averaging procedure, the norm of this state is given by
	\begin{equation}
		\braket{\varphi^{(BF)}}{\varphi^{(BF)}}_{\mathcal{H}_{BF}} = \braket{\varphi^{(B)}}{\varphi^{(BF)}}_{\mathcal{H}_B} = \frac{M^4}{16},
	\end{equation}
as would be expected by the procedure presented in section~\ref{fermionic-constraints}.
\section*{References}

\bibliographystyle{unsrt}

\bibliography{SUGRAProject}

\end{document}